\documentstyle  [preprint,tighten,aps,epsf]  {revtex}

\begin{document}
\input{epsf}
\draft

\newcommand   {\dzero}     {D\O}

\newcommand   {\qbar}      {\mbox{${\bar{q}}$}}
\newcommand   {\ubar}      {\mbox{${\bar{u}}$}}
\newcommand   {\dbar}      {\mbox{${\bar{d}}$}}
\newcommand   {\bbar}      {\mbox{${\bar{b}}$}}
\newcommand   {\tbar}      {\mbox{${\bar{t}}$}}
\newcommand   {\ppbar}     {\mbox{${p\bar{p}}$}}
\newcommand   {\qqbar}     {\mbox{${q\bar{q}}$}}
\newcommand   {\bbbar}     {\mbox{${b\bar{b}}$}}
\newcommand   {\ttbar}     {\mbox{${t\bar{t}}$}}
\newcommand   {\ee}        {\mbox{${e^+}{e^-}$}}
\newcommand   {\mm}        {\mbox{${\mu^+}{\mu^-}$}}
\newcommand   {\ep}        {\mbox{$e{\gamma}$}}
\newcommand   {\pp}        {\mbox{${\gamma}{\gamma}$}}

\newcommand   {\rar}       {\rightarrow}
\newcommand   {\rargap}    {\mbox{ $\rightarrow$ }}

\newcommand   {\vtb}       {\mbox{$V_{tb}$}}
\newcommand   {\vud}       {\mbox{$V_{ud}$}}
\newcommand   {\vma}       {\mbox{($V$--$A$)}}
\newcommand   {\vpa}       {\mbox{($V$+$A$)}}
\newcommand   {\shat}      {\mbox{$\hat{s}$}}
\newcommand   {\meb}       {\mbox{$m_{eb}$}}
\newcommand   {\costh}     {\mbox{$\cos\theta^*_e$}}

\newcommand   {\tev}       {\mbox{$\mbox{ TeV}$}}
\newcommand   {\gev}       {\mbox{$\mbox{ GeV}$}}
\newcommand   {\mev}       {\mbox{$\mbox{ MeV}$}}

\newcommand   {\bfppbar}   {\mbox{$\bf{p\bar{p}}$}}
\newcommand   {\bfqpqbar}  {\mbox{$\bf{q'\bar{q}}$}}
\newcommand   {\bfqqbar}   {\mbox{$\bf{q\bar{q}}$}}
\newcommand   {\bfrar}     {\bf{\rightarrow}}
\newcommand   {\bfrargap}  {\mbox{ $\bf{\rightarrow}$ }}
\newcommand   {\bftbbar}   {\mbox{$\bf{t\bar{b}}$}}
\newcommand   {\bftbbarq}  {\mbox{$\bf{t\bar{b}q}$}}
\newcommand   {\bftqbbar}  {\mbox{$\bf{tq\bar{b}}$}}
\newcommand   {\bftWbbar}  {\mbox{$\bf{tW\bar{b}}$}}
\newcommand   {\bftbar}    {\mbox{$\bf{\bar{t}}$}}
\newcommand   {\bftbarb}   {\mbox{$\bf{\bar{t}b}$}}
\newcommand   {\bfqpbarbbar} {\mbox{$\bf{\bar{q}'\bar{b}}$}}
\newcommand   {\bftbarqbar}{\mbox{$\bf{\bar{t}\bar{q}}$}}
\newcommand   {\bfqpbarg}  {\mbox{$\bf{\bar{q}'g}$}}
\newcommand   {\bftbarqbarb} {\mbox{$\bf{\bar{t}\bar{q}b}$}}
\newcommand   {\bftbarW}   {\mbox{$\bf{\bar{t}W}$}}


\preprint {INP-MSU-96-41/448, UCR/96-25}

\title    {Single top quarks at the Fermilab Tevatron}

\author   {A.P.~Heinson}

\address  {Department of Physics, University of California, Riverside,
           CA 92521, USA}

\author   {A.S.~Belyaev and E.E.~Boos}

\address  {Institute of Nuclear Physics, Moscow State University, RU-119899
           Moscow, Russia}

\maketitle

\begin{abstract}

We present a calculation of the single top quark cross section for
proton-antiproton interactions with $\sqrt{s}=1.8{\tev}$ at the Fermilab
Tevatron collider. We examine the effects of top mass, parton distribution
functions, QCD scale, and collision energy, on each of the component production
mechanisms, and study the kinematic distributions for standard model electroweak
production. At the upgraded Tevatron with $\sqrt{s}=2.0{\tev}$ and high
luminosity, it will be possible to test the nature of the $Wtb$ coupling using
single top production. We estimate the sensitivity to measure the single top
cross section, and thus to directly measure {\vtb} and the top quark partial
width. We show what happens to the {\vtb} measurement when an anomalous {\vpa}
component is added to the $Wtb$ coupling, and how the top quark polarization
affects the kinematic distributions.

\end{abstract}



\section*{Introduction}

Top quarks can be created via two independent production mechanisms in {\ppbar}
collisions. The primary mode, strong {\ttbar} pair production from a $gtt$
vertex, was used by the {\dzero} and CDF collaborations to establish the
existence of the top quark in March 1995 \cite{d01,cdf1}. The second mode is
electroweak production of a single top quark or antiquark from a $Wtb$ vertex.
This mechanism provides a sensitive probe for several standard model parameters.

The experimental value of the top quark mass is $175\pm{6}{\gev}$
\cite{grannis}, which is in good agreement with the value of $177\pm{7}
\mbox{ }_{-19}^{+16}{\gev}$ derived from electroweak measurements at LEP, SLC,
SPS, Tevatron, and neutrino scattering experiments \cite{ewmt}. Since the mass
is of the order of the electroweak symmetry breaking scale (vacuum expectation
value = 246{\gev}), it is a very promising place to look for deviations from the
standard model \cite{peccei}.

The coupling between a $W$~boson and top and bottom quarks has not yet been
studied directly. Top quark pair production is not the best process for probing
this $Wtb$ coupling or the Cabibbo-Kobayashi-Maskawa (CKM) matrix element {\vtb}
\cite{kobayashi} since the top quarks are produced from a $gtt$ vertex.
Information about $Wtb$ coupling from top quark decay is relatively inaccessible
in {\ttbar} pair production because the width of the decay into $Wb$ is
proportional to the branching fraction of $t{\rar}Wb$, which is close to unity
in the standard model (the top quark partial width $\Gamma(t{\rar}Wb)\propto
V_{tb}^2$, where $0.9989\leq|{\vtb}|\leq0.9993$ at the 90\% confidence level
\cite{pdg}). Single top quarks are produced at hadron colliders mainly from a
$Wtb$ vertex, and thereby provide a direct probe of the nature of the $Wtb$
coupling and of {\vtb}.

In hadron collisions, there are several partonic processes which produce single
top quarks in the final state. The $W$-gluon fusion mechanism
$q'g{\rar}tq{\bbar}$ has been studied in Refs.\ \cite{dawson,dicus,%
willenbrock,yuan2,moers,slabosp,parke,bordes,yuan3,bordes2,yuan3a,yuan3b,%
heinson}. The $W^*$ s-channel process $q'{\qbar}{\rar}t{\bbar}$ has been
examined in Refs.\ \cite{cortese,stelzer,pittau,smith,atwoodbar,li1,li2} for the
Tevatron. Use of single top quark production as a tool for studying the $Wtb$
coupling has been discussed in a number of papers for hadron colliders,
including the Tevatron \cite{yuan2,slabosp,heinson,yuan4,yuan5,yuan6,rizzo,%
mahlon}. Single top production has also been studied for {\ee}~colliders
\cite{yuan4,yuan5,yuan6,katuya,barger,raidal,panella,hagiwara,mele,boos,%
jikia1,booskuri,qian,ballestrero}, and for {\ep}~interactions \cite{jikia2,%
yehudai,pukhov}.

In this paper, we first present new results of consistent tree level cross
section calculations for two and three vertex subprocesses of single top quark
production in {\ppbar} collisions at $\sqrt{s}=1.8{\tev}$. We have chosen to use
this energy since it is where the Tevatron collider operated between 1992 and
1996, and this work is in support of a search of the data for single top
production. We then prepare the ground for single top physics at future high
luminosity Tevatron runs with $\sqrt{s}=2.0{\tev}$, by making new estimates of
the sensitivity to measure the top quark partial width, $Wtb$ coupling, and
{\vtb}, including an anomalous {\vpa} coupling. These studies are the first to
be performed using complete tree level matrix element calculations for all
possible processes, and do not use the effective $W$ approximation method
\cite{dawson}.

In section 1, we provide a comprehensive overview of the three separate single
top processes at the Tevatron, and their subprocesses. We describe our
computation techniques, and present and discuss the results of the calculations.
We have studied the cross section as a function of top quark mass, parton
distribution parametrization, choice of scale $Q^2$, and collider energy, and we
have evaluated lower and upper bounds on the single top cross section. In
section 2 of this paper, we investigate the kinematic distributions of single
top events, showing the separate contributions from the principal production
modes. In section 3, we look at the effects of a nonstandard coupling at the
$Wtb$ vertex, specifically the addition of an anomalous right-handed {\vpa}
contribution to the coupling. We estimate the sensitivity to determine {\vtb} by
measuring the single top cross section at future Tevatron runs as a function of
this possible right-handed coupling strength, and show how the polarization of
the system may be used to help distinguish different scenarios. We also use our
estimate of the cross section precision to predict the error on the top quark
partial width at the upgraded Tevatron. Finally, in section 4, we summarize our
results and draw conclusions.


\section{Single Top Quark Cross Section}

\subsection{Single top processes}

Within the standard model, there are three separate processes at a
proton-antiproton collider which result in a single top quark in the final
state. The list below shows that these processes in turn consist of several
subprocesses with two or three tree level vertices. The number of Feynman
diagrams for each subprocess is shown. Some diagrams have been omitted from the
total: (a) those with {\ttbar} pair production from a $gtt$ vertex and no
electroweak vertex (not single top); (b) those containing a photon, $Z$~boson or
Higgs (their contribution to the total cross section is extremely small); and
(c) diagrams with vertices containing off-diagonal CKM matrix elements. The
notation used is that $q$ is a light quark, and $X$ represents any additional
final state particles from the {\ppbar} interaction.
\begin{tabbing}
1.  ${\ppbar}$\=${\rargap}t{\bbar}$ \=$+X$\hspace{5mm}s-channe\=l
    $W^*$~boson\\*
      \>1.1      \>$q'{\qbar}{\rar}t{\bbar}$      \>1\\*
      \>1.2      \>$q'g      {\rar}t{\bbar}q$     \>2\\*
      \>1.3      \>$q'{\qbar}{\rar}t{\bbar}g$     \>4\\
2.  ${\ppbar}{\rargap}tq+X$\hspace{5mm}t- or u-channel $W$~boson\\*
      \>2.1      \>$q'b{\rar}tq$                  \>1\\*
      \>2.2      \>$q'g{\rar}tq{\bbar}$           \>2\\*
      \>2.3      \>$bg {\rar}tq{\qbar}'$          \>2\\*
      \>2.4      \>$q'b{\rar}tqg$                 \>4\\
3.  ${\ppbar}{\rargap}tW^-+X$\\*
      \>3.1      \>$bg {\rar}tW$                  \>2\\*
      \>3.2      \>$q{\qbar}{\rar}tW{\bbar}$      \>2\\*
      \>3.3      \>$gg      {\rar}tW{\bbar}$      \>5\\*
      \>3.4      \>$b{\bbar}{\rar}tW{\bbar}$      \>5\\*
      \>3.5      \>$qb {\rar}tWq$                 \>3\\*
      \>3.6      \>$bg {\rar}tWg$                 \>8
\end{tabbing}

It should be noted that there is some variation in the literature over the use
of the term ``$W$-gluon fusion". In some papers it refers only to subprocess
2.2 $q'g{\rar}tq{\bbar}$, in others to subprocesses 2.1 $q'b{\rar}tq$ and 2.2
$q'g{\rar}tq{\bbar}$ combined. Only one reference \cite{bordes2} includes all
four subprocesses 2.1--2.4 in the calculations, and the authors use the
term ``$W$-gluon fusion" to refer only to subprocess 2.2. In this paper, we
will also use the term to mean only subprocess 2.2.

The two subprocesses 1.2 and 2.2 (both $q'g{\rar}tq{\bbar}$), although
superficially similar, each contain two different Feynman diagrams which are
gauge invariant in separate pairs and do not need to be calculated together as
a group of four diagrams. One might be tempted to consider the four diagrams as
an independent set instead of as separate higher order corrections to the two
main processes, since an experimental search will be able to distinguish
between two-body and three-body final states. However, this is not acceptable
mathematically when calculating the cross sections, due to the definition of
the $b$ sea quarks in the parton distributions for the proton and antiproton.
This definition requires the subprocesses to be grouped as shown in the list
above.

\begin{figure}
\epsfxsize=6.5in \epsfbox{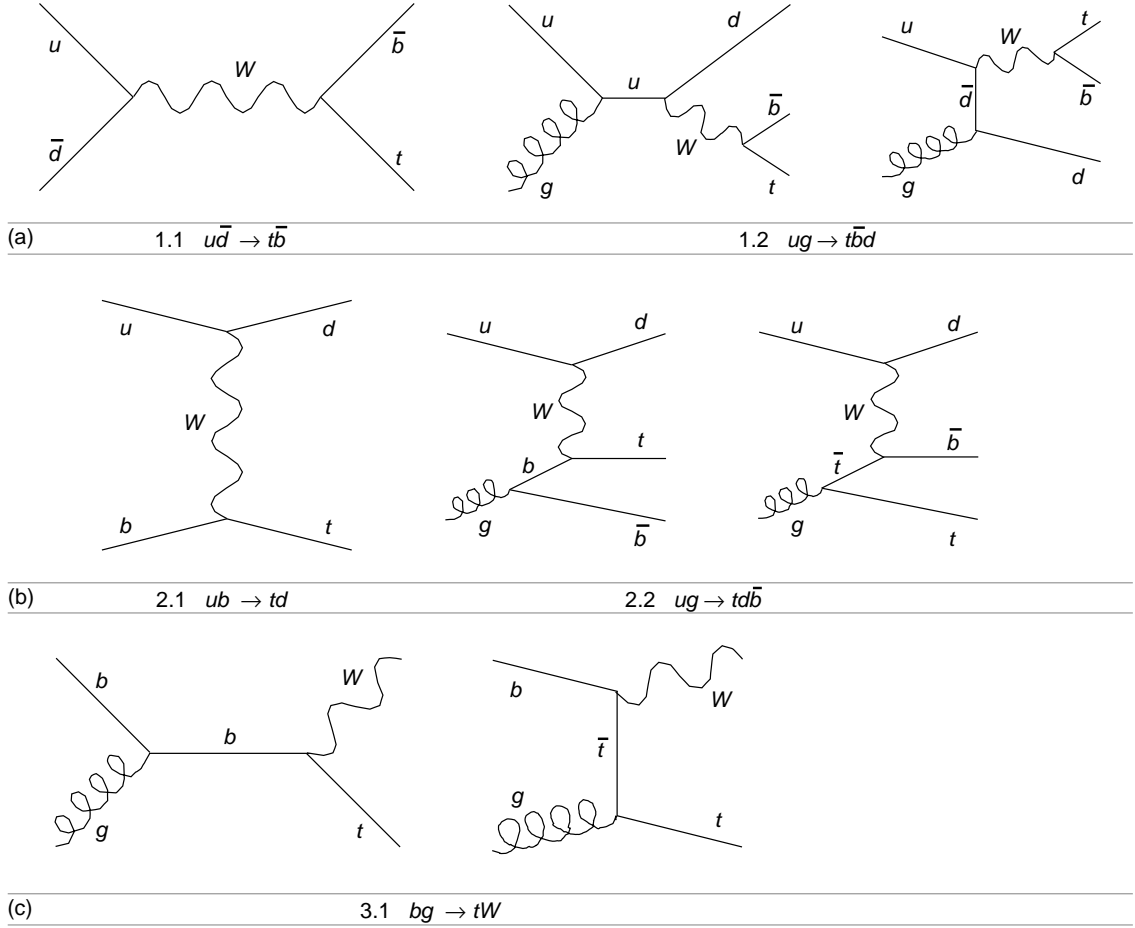}

\caption[fig1]{Representative Feynman diagrams for the three single top
quark production processes at the Fermilab Tevatron: (a) the $W^*$ s-channel
process ${\ppbar}{\rargap}t{\bbar}+X$; (b) the $W$ t- and u-channel process
${\ppbar}{\rargap}tq+X$, including subprocess 2.2, $W$-gluon fusion; and
(c) ${\ppbar}{\rargap}tW^-+X$.}

\label{feynman}
\end{figure}

We will now discuss the subprocesses we have included in our calculations. In
the following list, the first initial state particle is a parton from the
proton and the second one is a parton from the antiproton. Processes with an
initial state $c$ or $s$ quark, or with off-diagonal CKM matrix elements are
omitted from the list (and from plots) for simplicity, but have been included
in our calculation of the overall cross section and other numerical results.
\begin{tabbing}
1.  \=${\ppbar}{\rargap}$\= $t{\bbar}+X$\\*
 \>1.1 \>$u{\dbar}{\rar}t{\bbar},$\hspace{3mm}${\dbar}u{\rar}t{\bbar}$\\*
 \>1.2 \>$ug{\rar}t{\bbar}d,$\hspace{3mm}$gu{\rar}t{\bbar}d,$\hspace{3mm}
${\dbar}g{\rar}t{\bbar}{\ubar},$\hspace{3mm}$g{\dbar}{\rar}t{\bbar}{\ubar}$\\*

2.  ${\ppbar}{\rargap}tq+X$\\*
 \>2.1 \>$ub{\rar}td,$\hspace{3mm}$bu{\rar}td,$\hspace{3mm}
${\dbar}b{\rar}t{\ubar},$\hspace{3mm}$b{\dbar}{\rar}t{\ubar}$\\*
 \>2.2 \>$ug{\rar}td{\bbar},$\hspace{3mm}$gu{\rar}td{\bbar},$\hspace{3mm}
${\dbar}g{\rar}t{\ubar}{\bbar},$\hspace{3mm}$g{\dbar}{\rar}t{\ubar}{\bbar}$\\*

3.  ${\ppbar}{\rargap}tW^-+X$\\*
 \>3.1 \>$bg{\rar}tW,$\hspace{3mm}$gb{\rar}tW$\\*
 \>3.2 \>$u{\ubar}{\rar}tW{\bbar},$\hspace{3mm}${\ubar}u{\rar}tW{\bbar},
$\hspace{3mm}$d{\dbar}{\rar}tW{\bbar},$\hspace{3mm}${\dbar}d{\rar}tW{\bbar}$\\*
 \>3.3 \>$gg{\rar}tW{\bbar}$
\end{tabbing}
If the initial state parton is a $u$~quark from the proton, then contributions
from both valence and sea $u$~quarks are included in the calculations. This
also applies to $d$~quarks in the proton and to antiquarks in the antiproton.
Typical Feynman diagrams for these processes are shown in Fig.\ \ref{feynman}.

We have included in our calculations all the significant single top
subprocesses with two or three vertices, except those with a gluon in the final
state, which are significant, but which require a full next-to-leading order
calculation to be included properly. Subprocesses with an extra quark in the
final state, for instance $bg{\rar}tq{\qbar}'$ and $qb{\rar}tWq$, although they
have several Feynman diagrams, only contribute 1.5\% to the total
${\ppbar}{\rargap}tq+X$ cross section and 1\% to the ${\ppbar}{\rargap}tW+X$
rate, and have therefore been ignored. This also applies to the subprocess
$b{\bbar}{\rar}tW{\bbar}$, despite its having multiple Feynman diagrams,
including ones with electroweak {\ttbar} production.


\subsection{Calculation details}

We have calculated the production cross section for each of the single top
subprocesses mentioned in the previous section. We used the software package
{\sc c{\small{omp}}hep} \cite{comphep} to do the tree level symbolic
calculations and to generate optimized {\sc fortran} code for the squared
matrix elements. Version 2.0 of {\sc c{\small{omp}}hep} used the {\sc bases}
package \cite{bases} to integrate over all phase space using parton
distributions, and a {\sc c{\small{omp}}hep--bases} interface program to
generate the correct event kinematics, with smoothing of singular variables
\cite{kovalenko}. The Monte Carlo event generator {\sc spring} \cite{bases} was
used for each process in {\sc c{\small{omp}}hep}~2.0. {\sc
c{\small{omp}}hep}~3.0 has since replaced {\sc bases} and {\sc spring} with
{\sc vegas} \cite{vegas}, and we used this version as well. Events generated
were processed using {\sc pythia} \cite{pythia} via a custom interface, in
order to decay the $W$~boson for use in kinematic studies of the final state
particles.

For these calculations, we have utilized the {\sc cteq3m} \cite{cteq} and {\sc
mrs(a$^\prime$)} \cite{mrs} parton distributions. These two sets of
next-to-leading order structure functions both use the modified minimal
subtraction ($\overline{MS}$) renormalization scheme \cite{bardeen}. The newly
available parton distributions {\sc cteq4m} \cite{cteq4} and {\sc mrs(r)}
\cite{mrsr} are very similar to the distributions we have used, in the kinematic
region for single top quark production at the Tevatron.

We used the following standard model parameters in our calculations: $Z$~boson
mass $m_Z=91.19{\gev}$, $\sin^2{\theta_w}=0.225$, where $\theta_w$ is the weak
mixing angle, (giving the $W$~boson mass $m_W=m_Z\cos{\theta_w} = 80.28{\gev}$),
$b$~quark mass $m_b=5.0{\gev}$, $\alpha=1/128$, and CKM matrix elements
${\vud}=0.975$ and ${\vtb}=0.999$. All results have been obtained in two gauges,
the unitarity gauge and the 't~Hooft-Feynman gauge, as a check of calculations.
Differences between calculations in the two gauges are less than 0.1\%.

We have chosen to use $m_t^2$ as the QCD evolution parameter or scale $Q^2$
value, since a large scale is a natural choice for top quark production. A high
value such as $m_t^2$ is also a conservative choice that leads to lower cross
sections. As shown in Refs.\ \cite{parke} and \cite{yuan3a}, the leading order
single top cross section depends rather strongly on the choice of QCD scale for
values below $\sim$$(m_W/2)^2$, and if a very small value for $Q^2$ such as
$m_b^2$ were to have been chosen, then the calculated cross sections would have
been about four times larger than those obtained using $m_t^2$.

A typical $x$ value for single top quark processes is $\sim m_t/\sqrt{s}$
${\approx}$ $180/1800=0.1$, where $x$ is the fraction of the proton or
antiproton momentum carried by each initial state parton. At a scale
$Q^2=(180{\gev})^2$, the value of $\alpha_s$ is 0.102 from {\sc cteq3m} and
0.104 from {\sc mrs(a$^\prime$)}. ${\Lambda}_{QCD}$ for five quark flavors is
158.0{\mev} in both {\sc cteq3m} and {\sc mrs(a$^\prime$)}.


\subsection{Combining cross sections}

Care must be taken when combining some single top subprocesses in order to avoid
double counting. One cannot simply add up the separate cross sections to get the
total when there is a sea $b$ quark in the initial state. The {\sc cteq3m} and
{\sc mrs(a$^\prime$)} $b$ sea distributions are not measured experimentally, but
are obtained from the gluon distributions using the
Dokshitzer-Gribov-Lipatov-Altarelli-Parisi (DGLAP) evolution equations
\cite{dglap}. The $b$ sea distributions in the structure functions therefore
contain a mass singularity from the collinear divergence which occurs when the
gluon splits to an onshell $b{\bbar}$ pair. The subprocesses we are considering
which pertain to this situation are 2.1 $q'b{\rar}tq$, and 2.2
$q'g{\rar}tq{\bbar}$ ($W$-gluon fusion), where the initial state $b$~quark in
subprocess 2.1 is derived from the gluon sea in the antiproton. The correct way
to avoid this singularity would be to calculate the rate for subprocess 2.2
$q'g{\rar}tq{\bbar}$ with complete loop corrections, and then subprocess 2.1
with its $b$ sea contribution would be automatically included without the need
for extracting it from the parton distribution sets. However, since we are
making leading order calculations, we need another method. One technique
\cite{barnett,olness} for obtaining the cross section
$\sigma({\ppbar}{\rargap}tq+X)$ is to calculate the rates for subprocesses 2.1
and 2.2 and add them together, and then subtract the rate from the splitting
process $g{\rar}b{\bbar}$ convoluted with subprocess $q'b{\rar}tq$. We have
chosen to employ this method here. For instance, at $m_t=180{\gev}$ with
$Q^2=m_t^2$ and {\sc cteq3m} parton distributions, the naive cross section for
$q'b{\rar}tq$ is 0.75~pb, for $q'g{\rar}tq{\bbar}$ it is 0.29~pb, and the
splitting term is 0.54~pb, giving a total cross section from these two
subprocesses of 0.50~pb. The rate for ${\ppbar}{\rargap}tq+X$ is thus over 70\%
higher than the rate from the $W$-gluon fusion subprocess $q'g{\rar}tq{\bbar}$
alone.

A second subtlety \cite{tung} comes into play with this method for avoiding
double counting, when working with the two parton distributions being
considered: {\sc cteq3m} and {\sc mrs(a$^\prime$)}. The CTEQ collaboration has
chosen to evolve the $b$ sea distribution from $m_b$, whereas the MRS group
starts the evolution of its $b$ sea distribution at $2m_b$. The logarithmic
terms in the gluon splitting must be evaluated as $\ln(Q^2/m_b)$ with CTEQ
distributions, but as $\ln(Q^2/2m_b)$ for MRS distributions, to be consistent
with the respective definitions of the $b$~quark sea.


\subsection{Cross section versus top quark mass}

\begin{figure}
\epsfxsize=6.5in \epsfbox{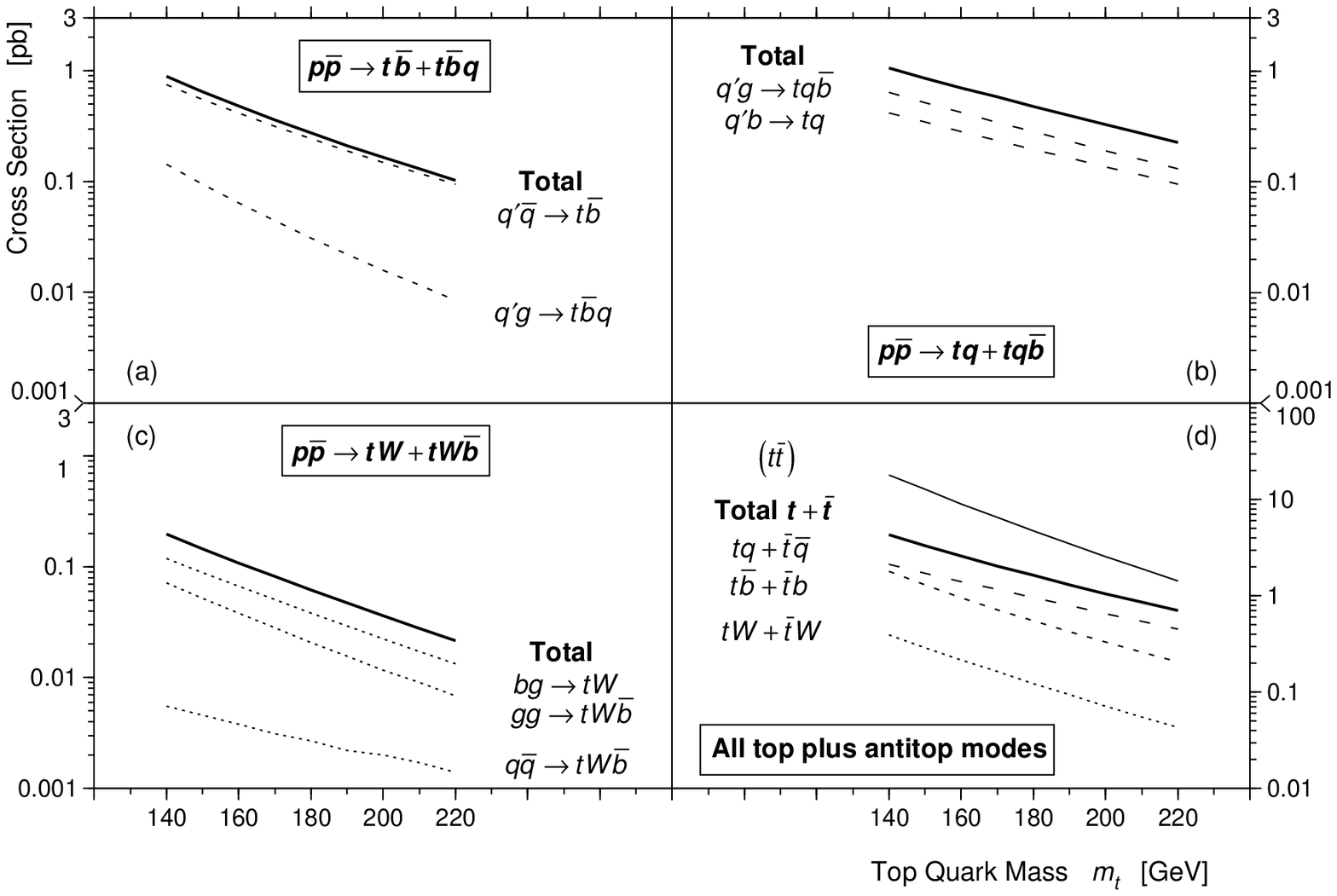}

\caption[fig2]{Single top quark cross sections at the Tevatron with
$\sqrt{s}=1.8$~TeV, versus top quark mass: (a) s-channel $W^*$ production
${\ppbar}{\rargap}t{\bbar}+t{\bbar}q$; (b) t- and u-channel production
${\ppbar}{\rargap}tq+tq{\bbar}$; (c) ${\ppbar}{\rargap}tW+tW{\bbar}$; and (d)
the total single top and antitop cross section ${\ppbar}{\rargap}t+{\tbar}+X$.
The resummed next-to-leading order ${\ttbar}$ cross section of Ref.\
\cite{berger} is shown as the uppermost line in (d), for comparison with single
top production (at leading order).}

\label{xsecmass}
\end{figure}

We show results for the cross sections of the three electroweak single top
processes and the totals in Fig.\ \ref{xsecmass} as a function of the top quark
mass, with $\sqrt{s}= 1.8{\tev}$. Figure\ \ref{xsecmass}(a) shows
${\ppbar}{\rargap}t{\bbar}+X$, (b) portrays the process ${\ppbar}{\rargap}tq+X$,
(c) is for the less important ${\ppbar}{\rargap}tW+X$ mode, and (d) shows the
totals for each of these three processes for $t$ and {\tbar} combined. Figure\
\ref{xsecmass}(d) also shows {\ttbar} pair production for comparison (upper
line), and it can be seen that when only one top quark is produced in the final
state, the cross sections decrease more slowly with increasing top quark mass
than when two heavy tops have to be created at once. The strong {\ttbar} cross
section illustrated is from the resummed next-to-leading order calculation of
Berger and Contopanagos \cite{berger}, who used the {\sc cteq3m} parton
distributions. The tree level single top cross sections are the average of our
calculations using {\sc cteq3m} and {\sc mrs(a$^\prime$)}.

The main contribution to electroweak single top production comes from
${\ppbar}{\rargap}tq+X$, the $W$~boson t- and u-channel mode, including
$W$-gluon fusion. The rate from this process (61\%, at $m_t=180{\gev}$)  is
nearly twice as large as that from ${\ppbar}{\rargap}t{\bbar}+X$ with a $W^*$ in
the s-channel (32\%). The contribution to the total cross section of the third
process ${\ppbar}{\rargap}tW+X$ is small (7\%). Of the dominant t- and u-channel
process, 41\% of the rate comes from $q'b{\rar}tq$ (after subtraction of the
splitting term), and 59\% from $W$-gluon fusion $q'g{\rar}tq{\bbar}$. Therefore,
$W$-gluon fusion forms 36\% of the total single top rate from all processes.

Because there are contributions from several single top processes, the total
cross section forms a significant fraction of the {\ttbar} pair production
rate. The single top and antitop cross section from {\ppbar} production at
$\sqrt{s}=1.8{\tev}$ is $0.92\times2=1.84$~pb for a top quark of mass 180{\gev}
and the {\sc cteq3m} parton distributions, and $0.84\times2=1.68$~pb using {\sc
mrs(a$^\prime$)}. Therefore, although the rate of single top production is
smaller than that from {\ttbar} pair production (e.g. $4.71^{+0.07}_{-0.35}$~pb
at 180{\gev} \cite{berger}) for all top quark masses considered here, it is
large enough to be extremely interesting for study at the Tevatron.

Recent calculations show that higher order corrections to the leading order
single top cross sections presented here are large. For instance, Ref.\
\cite{bordes2} shows that the $K$ factor for ${\ppbar}{\rargap}tq+X$ is
$\sim$1.45 for $m_t=180{\gev}$ at $\sqrt{s}=1.8{\tev}$ with {\sc cteq2d} parton
distributions \cite{cteq2} and $Q^2=m_t^2$. Reference\ \cite{smith} contains a
similar higher order calculation for ${\ppbar}{\rargap}t{\bbar}+X$, and finds
that for $m_t=175{\gev}$ and $\sqrt{s}=2.0{\tev}$, the $K$ factor is also 1.45,
using the {\sc cteq3m} parton distributions and $Q^2=m_W^2$.


\subsection{Contributions to the single top cross section}

Table~\ref{table1} presents values of various partonic subprocess cross sections
for a top quark of mass 180{\gev}, and for the two parton distributions
discussed previously. Subprocesses with an initial state strange or charm sea
quark contribute 1.9\% to the total ${\ppbar}{\rargap}t{\bbar}+X$ cross section,
and 6.1\% to the total ${\ppbar}{\rargap}tq+X$ rate. Off-diagonal CKM matrix
element subprocesses (not including initial state $s$ and $c$ sea quark
subprocesses) contribute 0.3\% to ${\ppbar}{\rargap}t{\bbar}+X$ and 5.0\% to
${\ppbar}{\rargap}tq+X$. All these other modes contribute $<0.5\%$ to
${\ppbar}{\rargap}tW+X$ production. Off-diagonal CKM subprocesses and initial
state $s$ and $c$ sea quark subprocesses are included in our calculation of the
total single top cross section, but are not included in our plots for
simplicity, because of the calculation technique used.

For single top modes like $W^*$ s-channel production with only light partons in
the initial state, including both valence and sea quarks, the cross sections
calculated with {\sc mrs(a$^\prime$)} are 2.4\% lower than those calculated with
{\sc cteq3m}. When there is a gluon in the initial state, for instance $W$-gluon
fusion, then the {\sc mrs(a$^\prime$)} cross sections are 5.7\% lower than the
{\sc cteq3m} ones. For reactions with a sea $b$~quark in the initial state, the
cross sections calculated with {\sc mrs(a$^\prime$)} are 17\% smaller.

Table~\ref{table2} shows the resulting single top cross sections as a function
of top quark mass. The central value numbers are the mean of the values
calculated using {\sc cteq3m} and {\sc mrs(a$^\prime$)}. The upper and lower
bounds come from combining half the difference between the calculations using
the two parton distributions with the errors from the choice of QCD scale, as
discussed in the next section. The correlation of the errors is correctly
accounted for by adding the $Q^2$ errors to each subprocess separately, before
adding them in quadrature with the uncertainty due to the choice of structure
function. For a top quark of mass 180{\gev}, the total single top plus antitop
cross section is $1.76_{-0.18}^{+0.26}$~pb.


\subsection{Cross section versus scale}

\begin{figure}
\epsfxsize=6.5in \epsfbox{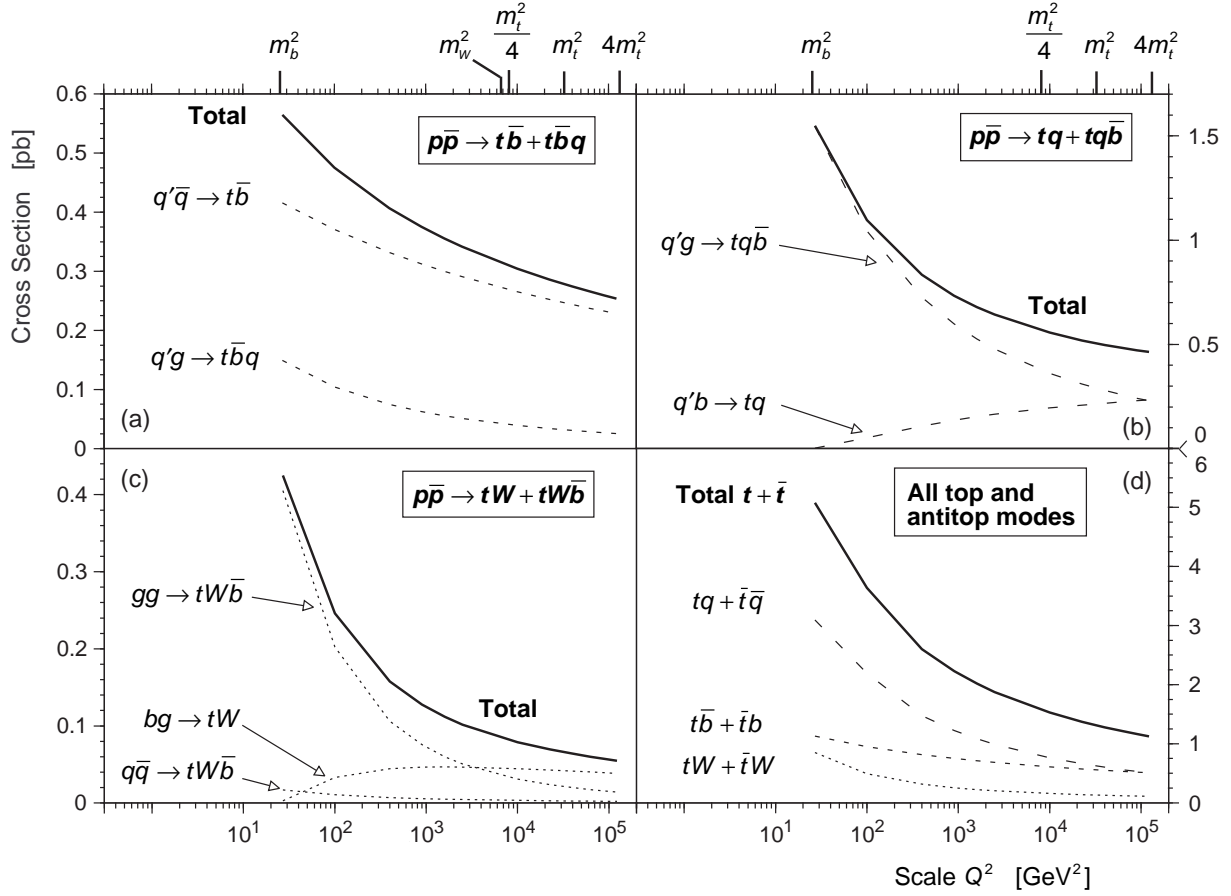}

\caption[fig3]{Single top quark cross sections ($m_t=180$~GeV,
$\sqrt{s}=1.8$~TeV) versus QCD evolution scale $Q^2$ for: (a) s-channel $W^*$
production ${\ppbar}{\rargap}t{\bbar}+t{\bbar}q$; (b) t- and u-channel
production ${\ppbar}{\rargap}tq+tq{\bbar}$; (c) ${\ppbar}{\rargap}tW+tW{\bbar}$;
and (d) the summed single top and antitop cross section
${\ppbar}{\rargap}t+{\tbar}+X$.}

\label{scale}
\end{figure}

We have examined the effect of the choice of QCD evolution parameter $Q^2$ on
the various single top subprocesses. The results are shown in Fig.\ \ref{scale}
for a top quark of mass 180{\gev} and the {\sc cteq3m} parton distributions at
$\sqrt{s}=1.8{\tev}$. Figure\ \ref{scale}(a) shows the scale dependence for the
$W^{*}$ s-channel process ${\ppbar}{\rargap}t{\bbar}+X$, which is the least
dependent of the various single top processes on the choice of scale. Figure\
\ref{scale}(b) is for the t- and u-channel processes $q'g{\rar}tq{\bbar}$
($W$-gluon fusion) and $q'b{\rar}tq$. The $W$-gluon fusion cross section falls
rapidly as the calculation scale increases, whereas the subprocess $q'b{\rar}tq$
goes up as $Q^2$ is raised. When these subprocesses are combined, the two
effects partially cancel. The $q'b{\rar}tq$ subprocess is shown with the
$g{\rar}b{\bbar}$ splitting term already subtracted. The minor single top
process ${\ppbar}{\rargap}tW+X$ is shown in Fig.\ \ref{scale}(c) with its
various contributing subprocesses, which again have differing dependences on
$Q^2$ that partially cancel in the sum. Finally, Fig.\ \ref{scale}(d) shows each
single top process summed for $t$ and ${\tbar}$, and the total single top
production on the same axis for comparison.

Leading order cross sections show more sensitivity to the choice of scale than
higher order calculations. However, one still needs to choose a scale at which
to perform the calculations. From an intuitive perspective, it does not make
sense to consider top quark production as occuring at the almost zero mass $b$
quark scale, although it would be mathematically consistent. Therefore, for our
calculations, we have chosen the central value of the scale to be $Q^2=m_t^2$,
and when estimating the uncertainty due to the choice of scale, we have
restricted the region of interest to lie between $(m_t/2)^2$ and $(2m_t)^2$, as
shown on the upper axes of Fig.\ \ref{scale}. The resulting errors on the main
contributions to the cross section at $m_t=180{\gev}$ are: $^{+9}_{-11}\%$ for
${\ppbar}{\rargap}t{\bbar}+X$; $^{-8}_{+9}\%$ for $q'b{\rar}tq$ and
$^{+32}_{-20}\%$ for $q'g{\rar}tq{\bbar}$, which combine to give $^{+15}_{-8}\%$
for ${\ppbar}{\rargap}tq+X$; and $^{+29}_{-14}\%$ for ${\ppbar}{\rargap}tW+X$.
The $q'b{\rar}tq$ and $q'g{\rar}tq{\bbar}$ errors largely cancel because the
contributions to the errors from choice of $Q^2$ are 100\% anticorrelated. The
$Q^2$ scale error dominates the total errors on the cross sections given in
Table~\ref{table2}.

\begin{figure}
\epsfxsize=6.5in \epsfbox{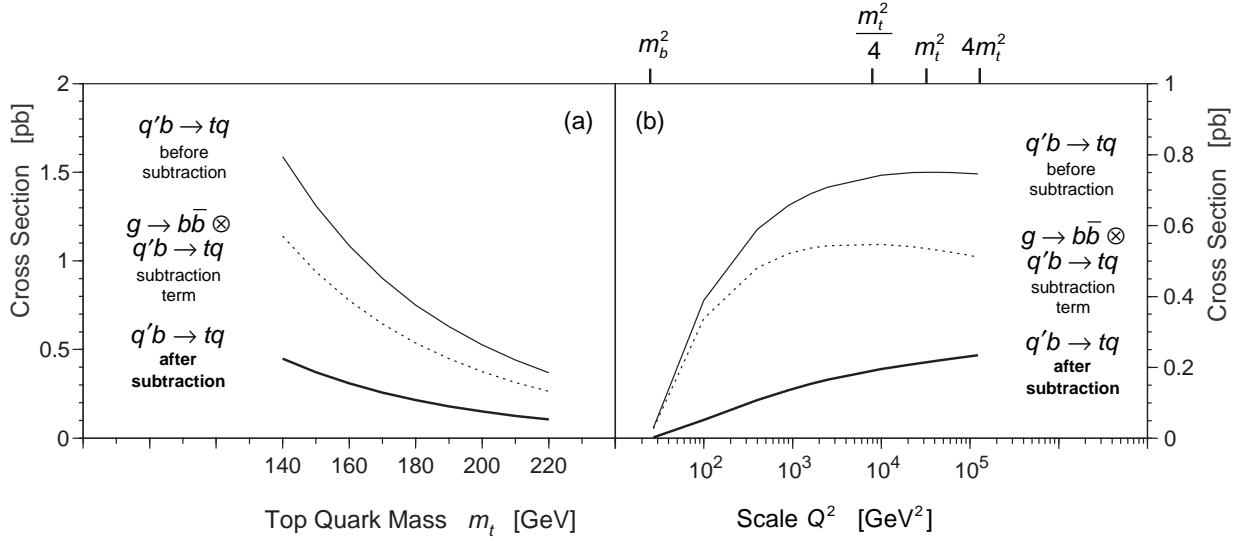}

\caption[fig4]{Single top produced together with a light quark, $q'b{\rar}tq$,
from an initial state sea $b$ quark, showing the cross section before and after
subtraction of the gluon splitting term, as a function of: (a) top quark mass
(with $Q^2=m_t^2$); and (b) scale $Q^2$ (with $m_t=180$~GeV).}

\label{split}
\end{figure}

Combining the subprocesses $q'b{\rar}tq$ and $q'g{\rar}tq{\bbar}$ by
subtracting the gluon splitting term to avoid double counting, as discussed
earlier, is a procedure that is sensitive to the choice of evolution parameter
$Q^2$. Figure\ \ref{split} shows the subprocess $q'b{\rar}tq$ before and after
subtraction, as a function of (a) the top quark mass, and (b) the scale $Q^2$.
It can be seen however, that provided the scale remains in the region around
$m_t^2$, then the sensitivity is less than that seen for the $W$-gluon fusion
subprocess $q'g{\rar}tq{\bbar}$ in Fig.\ \ref{scale}(b).


\subsection{Cross section versus collider energy}

\begin{figure}
\epsfxsize=6.5in \epsfbox{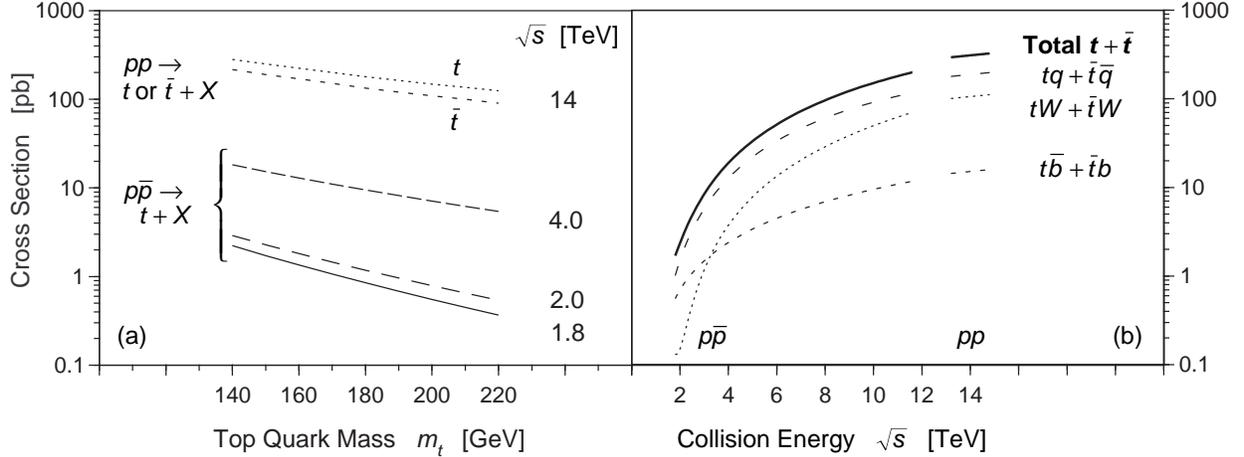}

\caption[fig5]{Single top quark cross section plotted (a) versus top quark
mass, at four production energies: the Fermilab Tevatron at $\sqrt{s}=1.8$~TeV;
the upgraded Tevatron at 2.0~TeV; the proposed TeV* collider at 4.0~TeV; and the
CERN $pp$ Large Hadron Collider at 14~TeV. Plot (b) shows the cross section
versus collider energy (with $m_t=180$~GeV), for each of the single top
production mechanisms. The values in (b) up to 12~TeV are for ${\ppbar}$
production, whereas the results at 14~TeV are for $pp$ collisions.}

\label{xsecenergy}
\end{figure}

We have calculated the single top quark cross section as a function of
production energy $\sqrt{s}$. Figure~\ref{xsecenergy}(a) shows the cross section
versus top quark mass for four collision energies: (i) the current Tevatron
energy 1.8{\tev}; (ii) the Tevatron energy for the next run in 1999, 2.0{\tev};
(iii) the energy of a possible Tevatron upgrade, 4.0{\tev}; and (iv) the energy
of the Large Hadron Collider (LHC) at CERN in 2005, 14{\tev}. The three Tevatron
cross sections are for {\ppbar} collisions, whereas the LHC cross sections are
calculated for $pp$ collisions. Despite the $\sim$150$\times$ increase in cross
sections at the LHC, it will still be rather difficult to study single top quark
production there, since the backgrounds will be much larger, and the signal will
be harder to identify, because the jet produced at the same time as the top
quark in $W$-gluon fusion for instance, will be further forward in
pseudorapidity $\eta$, where $\eta=\ln\tan(\theta/2)$ and $\theta$ is the polar
angle between the jet and the proton beamline. Peaks in the accompanying jet
distribution at the LHC will occur at $\eta=\pm2.5$ (c.f. $\eta$ peaks at
$\pm$1.5 when $\sqrt{s}=1.8{\tev}$).

At $m_t=180{\gev}$, the cross section for single top quark production is
0.85~pb at 1.8{\tev}, 1.4~pb at 2.0{\tev}, 9.4~pb at 4.0{\tev} and for $pp$
collisions at 14{\tev}, 179~pb. For 180{\gev} {\tbar}~antiquarks, the cross
sections are the same as for $t$ quarks at the Tevatron, but only 133~pb at the
LHC (26\% lower), because there are no valence antiquarks in the initial state.
These calculations were done using the {\sc cteq3m} parton distributions with
$Q^2=m_t^2$, and no contributions from initial state $s$ or $c$ quarks or
off-diagonal CKM matrix element terms are included.

The relative contributions to the total single top cross section from each of
the significant processes is not the same at all production energies. For a top
quark of mass 180{\gev}, Fig.\ \ref{xsecenergy}(b) shows the single top plus
antitop cross section versus production energy at the Tevatron for each
component of the signal separately. It can be seen that the $W^*$ s-channel
process ${\ppbar}{\rargap}t{\bbar}+{\tbar}b+X$ is much less sensitive to the
change in available energy than the other processes, which increase rapidly in
rate as the initial state energy goes up. At $\sqrt{s}=1.8{\tev}$, the $W^*$
process forms 32\% of the total single top signal, at 2.0{\tev} it provides 29\%
of the cross section, and by 4.0{\tev} it contributes only 13\%. The
${\ppbar}{\rargap}t{\bbar}+{\tbar}b+X$ process behaves in this manner because it
is an s-channel process and its contribution to the total cross section comes
from the {\shat} threshold phase space region, which is independent of energy.
The reason why the total cross section increases with energy is that at
higher energies, regions of smaller $x$ in the proton structure functions are
probed, and this is where the parton distributions are larger.

The contribution from ${\ppbar}{\rargap}tW+X$ to the total single top cross
section increases from 7\% at $\sqrt{s}=1.8{\tev}$ through 9\% at 2.0{\tev}, to
20\% at 4.0{\tev}. At the LHC, $pp{\rar}tW+X$ will contribute 30\% of the
single top rate and 40\% of the antitop rate, and could therefore be an
important production mode in the future. On the other hand, at the LHC, the
s-channel process $pp{\rar}t{\bbar}+{\tbar}b+X$will fall to only 5\% of
the total single top rate, and will become experimentally inaccessible.


\subsection{A closer look at $W$-gluon fusion}

\begin{figure}
\epsfxsize=6.5in \epsfbox{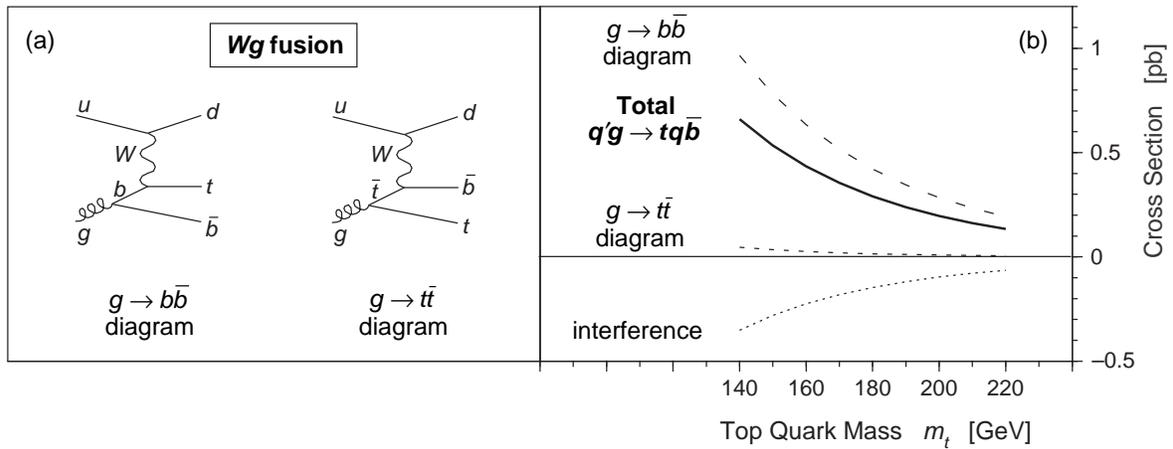}

\caption[fig6]{(a) Feynman diagrams for $W$-gluon fusion ($q'g{\rar}tq{\bbar}$).
(b) $W$-gluon fusion cross section versus top quark mass, showing the
contributions from each of the Feynman diagrams, and the large destructive
interference between the two processes.}

\label{xsecfusion}
\end{figure}

We have analyzed the contributions to the production rate from the two Feynman
diagrams which form $W$-gluon fusion, $q'g{\rar}tq{\bbar}$, shown in Fig.\
\ref{xsecfusion}(a). There is no interference between the $W$-gluon fusion
diagrams and the two nonfusion diagrams of subprocess $q'g{\rar}t{\bbar}q$
(shown in Fig.\ \ref{feynman}(a) as subprocess 1.2), because the final state
$t$~quark and {\bbar}~antiquark have a different color structure. For the
nonfusion diagrams, the $t$ and {\bbar} are from a $W$ decay and so are in a
color singlet state, whereas for the fusion diagrams the $t$ and {\bbar} come
from a gluon and so are in a color octet state.

The contribution to the total production rate of $W$-gluon fusion from the
Feynman diagram where the gluon produces a {\ttbar} pair is very small, at about
5\%. However, this diagram interferes destructively with the main $W$-gluon
fusion diagram where $g{\rar}{\bbbar}$. The destructive interference reduces
the total rate for $W$-gluon fusion by 34\%. We present the cross section versus
top quark mass for the two diagrams of $W$-gluon fusion separately, and show
the interference and net result, in Fig.\ \ref{xsecfusion}(b).


\subsection{More on ${\ppbar}{\rargap}tW+X$}

We have considered two related $2{\rar}3$ body processes in addition to the
process $bg{\rar}tW$. These are $q{\qbar}{\rar}tW{\bbar}$ and
$gg{\rar}tW{\bbar}$. We looked at these processes because in {\ee} and {\pp}
colliders, single top quark processes with $tW{\bbar}$ in the final state are
important. However, we found that at the Tevatron these processes are not very
significant. The interactions {\ee}, {\pp}, {\qqbar}, and $gg{\rar}tW{\bbar}$
all include diagrams with {\ttbar} pair production and subsequent decay of the
{\tbar} into $W{\bbar}$, as well as many additional diagrams with just single
top quark production. One needs to remove the contribution to the cross section
from the invariant mass region $m_t=m_{Wb}$ around the top quark pole in order
to study the $Wtb$ vertex in single top production. The remaining contributions
in {\ee} and {\pp} collisions are large enough (at 10~fb which is 17\% of the
total $tW{\bbar}$ cross section for {\ee} collisions at $\sqrt{s}=2{\tev}$ for
example) to be sensitive to the coupling structure, but in {\qqbar} and $gg$
collisions almost the entire cross section comes from the {\ttbar} diagrams, and
the remaining single top quark contribution at 29~fb, is only $\sim$0.8\% of the
total $tW{\bbar}$ rate of 3.5~pb.


\section{Kinematic Distributions}

In order to understand in more detail the properties of single top quark
production, we present in this section several experimentally interesting
kinematic distributions. These are shown for top production only
(not {\tbar}) to make the presentation clear. Distributions for antitop are the
same as those for top in transverse momentum, but are mirror images in
pseudorapidity. If the sign of the $W$~boson charge can be measured using its
leptonic decay mode, then it will be possible to study the properties of top
quarks and {\tbar}~antiquarks separately. All plots are for a top quark of mass
180{\gev} and have been calculated using the {\sc cteq3m} parton distributions
at $\sqrt{s}=1.8{\tev}$.

The top quark decays to a $W^+$~boson and a $b$~quark, and we consider here only
subsequent leptonic decays of the $W$ to a positron and neutrino, as this signal
should be easier to find experimentally than channels with hadronic decay of the
$W$~boson. The branching fraction $B$ for this decay mode is $\frac{1}{9}$. The
signature for a single top quark event is therefore a central, isolated, high
$p_T$ lepton), large missing transverse momentum from the neutrino, and at least
two jets, where one of the jets comes from the hadronization of the $b$~quark
from the decay of the top quark. All single top events therefore have one
potentially identifiable $b$~jet, and of the experimentally accessible
production modes at the Tevatron (${\ppbar}{\rargap}t{\bbar}+X$, $tq+X$),
$\sim$71\% of them have a {\bbar}~jet as well.


\subsection{Transverse momentum}

\begin{figure}
\epsfxsize=6.5in \epsfbox{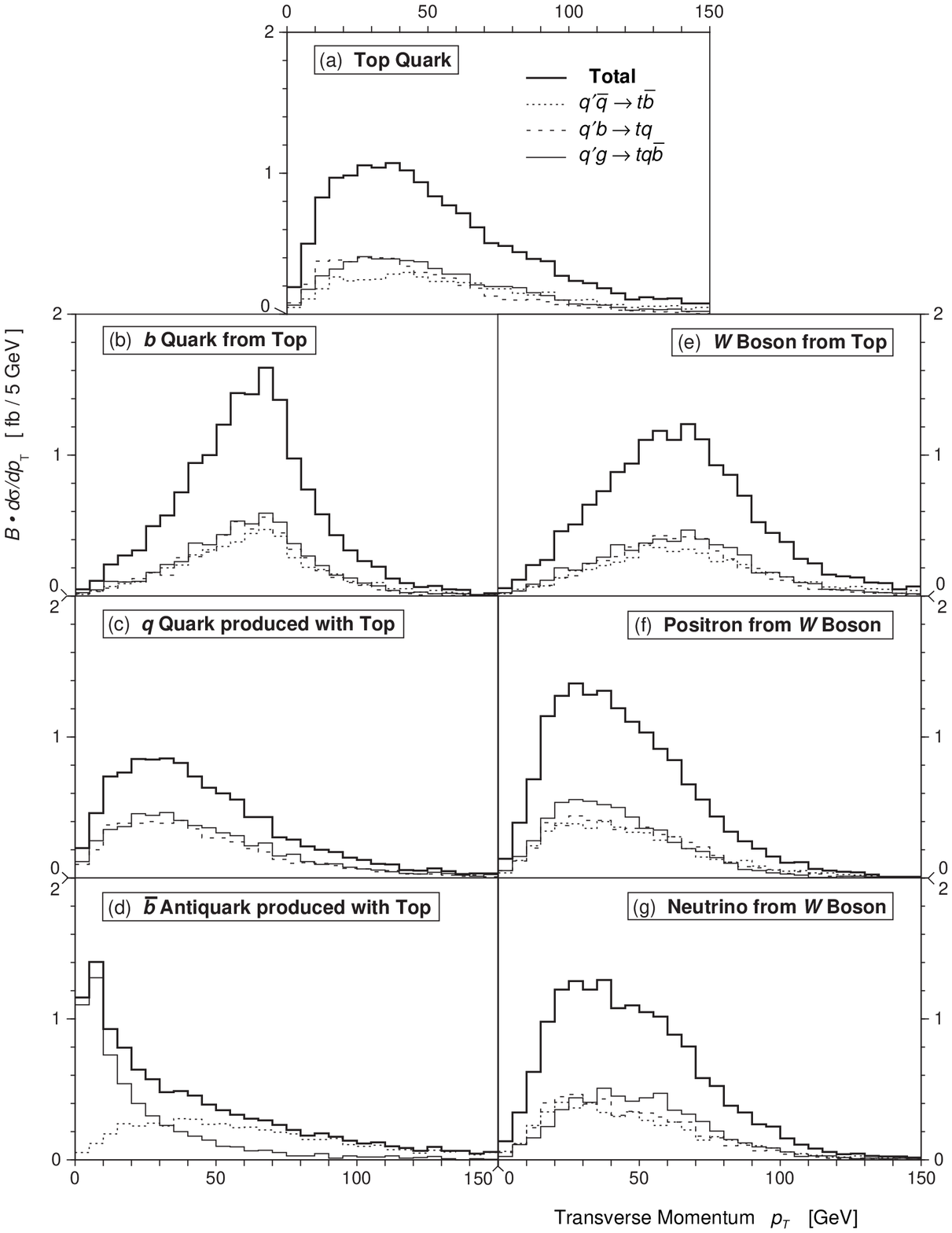}
 
\caption[fig7]{Single top transverse momentum distributions (with $m_t=180$~GeV)
for: (a) the top quark; (b) the $b$ quark from the decay of the top; (c) the
light $q$ quark produced with top in t- and u-channel processes; (d) the
${\bbar}$ antiquark produced with top in the s-channel $W^*$ process, and in
$W$-gluon fusion; (e) the $W$ boson from the top decay; (f) the positron from
the $W$ decay; and (g) the neutrino also from the decay of the $W$ boson.}
 
\label{transmom}
\end{figure}

Figure\ \ref{transmom} shows the branching fraction times differential cross
section $B \cdot d{\sigma}/dp_T$ versus transverse momentum $p_T$ of the final
state partons in single top production, and their decay products. In each plot,
the short-dashed line is for $W^*$ production $q'{\qbar}{\rar}t{\bbar}$, the
longer-dashed line for the two-body t-channel process $q'b{\rar}tq$, and the
narrow solid line for $W$-gluon fusion $q'g{\rar}tq{\bbar}$. The wide solid line
is the sum of these three processes. Plot (a) shows the transverse momentum
distributions of the top quark from each single top process. The mean of these
distributions is 51{\gev}. Despite its very high mass, the top quark is not
produced at rest, but carries considerable transverse momentum in all three
production modes. When the top decays, it produces a $b$~quark, whose $p_T$
distribution is shown in plot (b). The mean $p_T$ here is 62{\gev}. Plot (c) is
for the light quark produced with top in the t-channel processes ($\langle p_T
\rangle = 43{\gev}$), and (d) is for the {\bbar}~antiquark often produced with
top. Here the {\bbar} from $W^*$ single top production has $\langle p_T \rangle
= 59{\gev}$, whereas the {\bbar} in $W$-gluon fusion is much softer, with
$\langle p_T \rangle = 25{\gev}$. The low $p_T$ will make this jet much more
difficult to reconstruct. When the top quark decays, it produces a $W$~boson,
whose $p_T$ is shown in Fig.\ \ref{transmom}(e) ($\langle p_T \rangle =
65{\gev}$). The $W$ decays to a positron, (shown in (f)) with mean $p_T$ of
45{\gev} and a neutrino (in (g), 48{\gev}).


\subsection{Pseudorapidity}

\begin{figure}
 
\caption[fig8]{Single top pseudorapidity distributions (with $m_t=180$~GeV) for:
(a) the top quark; (b) the $b$ quark from the decay of the top; (c) the light
$q$ quark produced with top in t- and u-channel processes; (d) the ${\bbar}$
antiquark produced with top in the s-channel $W^*$ process, and in $W$-gluon
fusion; (e) the $W$ boson from the top decay; (f) the positron from the $W$
decay; and (g) the neutrino also from the decay of the $W$ boson.}
 
\epsfxsize=6.5in \epsfbox{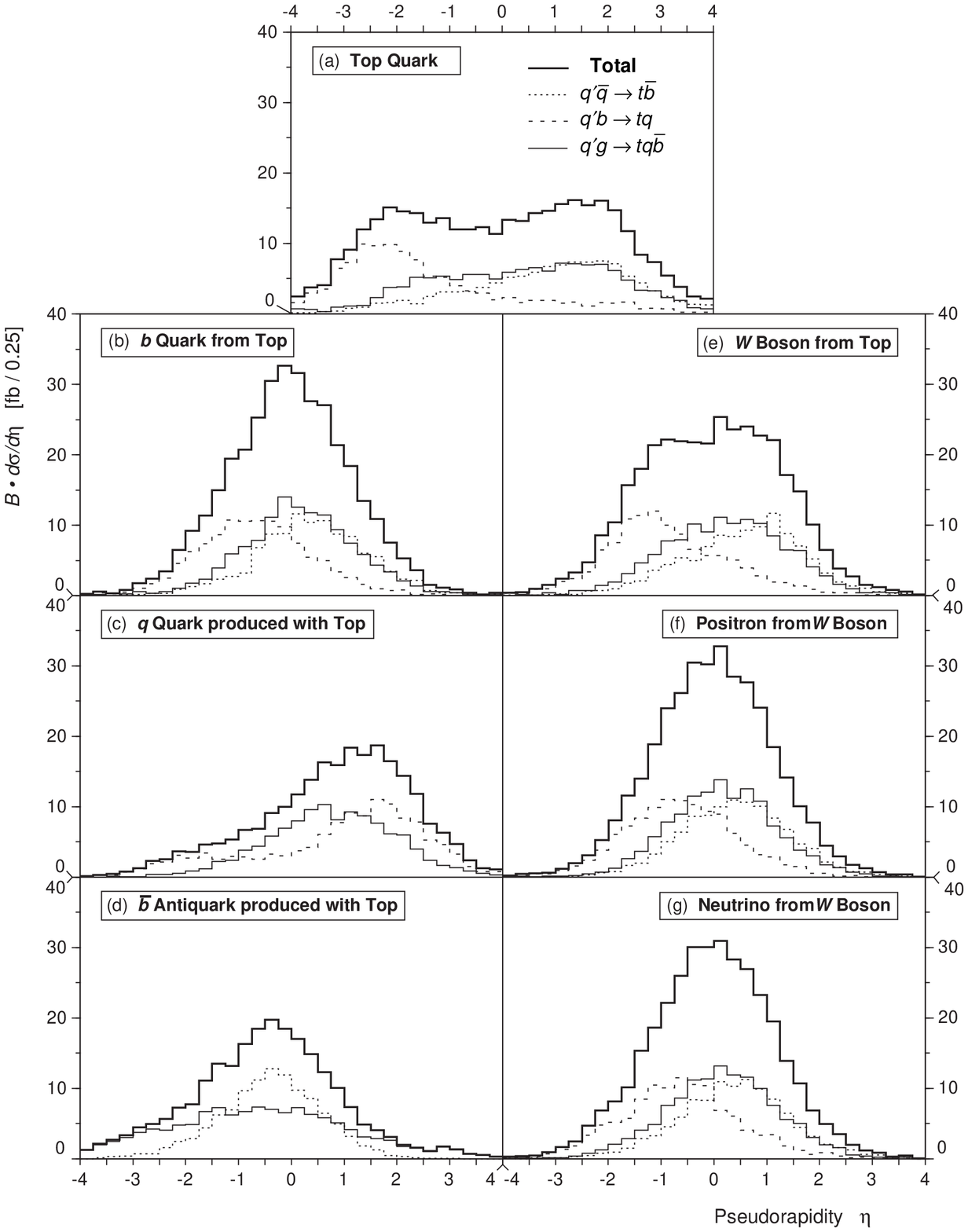}

\label{pseudorap}
\end{figure}

Figure\ \ref{pseudorap} shows the branching fraction times differential cross
section $B \cdot d{\sigma}/d\eta$ versus pseudorapidity $\eta$ of the final
state partons from single top production and their decay products. Plot (a) is
for the top quark itself, where one can see that the pseudorapidity
distributions are rather broad, and that the contributing production modes have
very different kinematics from each other. Both the $W$-gluon fusion and $W^*$
modes produce top quarks more in the forward or $+\eta$ direction than backwards
(with the distributions peaked at $\eta \sim 1.7$) whereas the two body
t-channel process $q'b{\rar}tq$ produces mainly backwards traveling top quarks,
with the peak at $\eta \sim -2.3$. This distribution is also narrower than the
other two. We see next that the decay products from the top are produced much
more centrally. Plot (b) shows the $b$~quark pseudorapidities. The distribution
for the $b$ from top decay in $W$-gluon fusion is peaked at $\eta \sim 0.1$, and
the $b$ from top in $W^*$ production at $\sim 0.2$. The $b$ from top in
$q'b{\rar}tq$ is still produced somewhat backwards, with a peak at $\sim-$0.8
reflecting the direction of its parent. We would like to note that the $\eta$
distribution of the $b$~quark from the top decay in $W$-gluon fusion is in
agreement with that seen by C.-P.~Yuan using the {\sc onetop} generator
\cite{yuan2}, but is rather different from the distribution for $W$-gluon fusion
shown in the TeV-2000 study of $WH$, $H{\rar}b{\bbar}$ \cite{tev2000a} (with
single top as a background), where the {\sc herwig} generator \cite{herwig} was
used for this type of single top. H{\sc erwig} seems to produce $b$'s in a
symmetric peak in the region $1<|\eta|<5$. This difference is not understood.

One of the striking features of $W$-gluon fusion is the forward direction in
which the light quark is produced \cite{yuan2}. This can be seen in plot (c),
where the light quark from $q'g{\rar}tq{\bbar}$ has a broad distribution, peaked
at $\sim 0.7$. The effect is seen more emphatically in the two body t-channel
mode where the peak occurs around 1.7, resulting in the summed distribution
peaking at $\eta \sim 1.5$. The pseudorapidity distributions of the
{\bbar}~antiquark produced together with top in 71\% of single top events are
shown in plot (d). Both distributions peak at $\eta \sim -0.4$; the soft {\bbar}
from $W$-gluon fusion has a rather broad spread in pseudorapidity, whereas the
much harder {\bbar} from $W^*$ production is produced in a narrower
pseudorapidity peak. The $\eta$ distributions of the $W$~boson from the decay of
the top quark, shown in plot (e), are peaked at $\sim 0.3$ for $W$-gluon fusion,
at $\sim 1.1$ for $W^*$ production, and at $\sim-$1.2 for the $q'b{\rar}tq$
mode, echoing the directions of their respective parent top quarks. The positron
(f) and neutrino (g) distributions are more central versions of their parent
$W$~bosons.


\section{$W{\lowercase{tb}}$ Coupling and $V_{\lowercase{tb}}$}

Since the top quark is rather heavy, we expect that new physics might be
revealed at the scale of its mass. Many variants of nonstandard physics relating
to this subject have been considered in the literature. Possible anomalous
gluon--top~quark couplings are discussed in Refs.\ \cite{deshpande,atwood,%
atwood2,huang,cheung,jberger,tait}. Contact terms and new strong dynamics
involving the top quark have been studied in \cite{hill,hill2,eichten,lsmith,%
datta,simmons}. The $Ztt$ coupling will be inaccessible until a high energy
{\ee} or {\mm} collider is in operation. Studies of the $Wtb$ coupling however,
will be possible before then using single top production at the Tevatron.

In this section we examine the effects on single top quark production and on its
decay kinematics of a deviation in the $Wtb$ coupling from the standard model
structure, and we consider how this will affect a measurement of the CKM matrix
element {\vtb}. In the standard model, the $Wtb$ coupling is proportional to
{\vtb} and has {\vma} structure. As explained in the introduction, the cross
section for single top quarks includes the $Wtb$ coupling directly, in contrast
to {\ttbar} pair production. Therefore, single top production provides a unique
opportunity to study the $Wtb$ structure and to measure {\vtb}. Experimental
studies of this type are among the main goals of single top physics. Because
high statistics will be required to make sensitive measurements, all the results
given in the remaining subsections of this paper are for single top events
produced in Runs~2 or 3 of the Tevatron; that is, from 1999 onwards, with a
collision energy of $\sqrt{s}=2.0{\tev}$.


\subsection{Anomalous {\vpa} coupling}

As an example of a deviation from the standard model $Wtb$ coupling, we
introduce an additional contribution from a nonstandard {\vpa} structure with an
arbitrary parameter $A_r$, where the subscript $r$ refers to the right-handed
current it represents. In the unitarity gauge, the $Wtb$ coupling is given by:
\[
     \Gamma = \frac{eV_{tb}}{2\sqrt{2}\sin\theta_w}
                  \left[\gamma_{\mu}\left(1-\gamma_5\right)
                  + A_r\gamma_{\mu}\left(1+\gamma_5\right)\right]
\]
\noindent
where $e$ is the positron electric charge, $\sin\theta_w=0.474$, and
$\gamma_{\mu}$ and $\gamma_5$ are Dirac matrices.

\begin{figure}
\leavevmode
\centering
\epsfbox{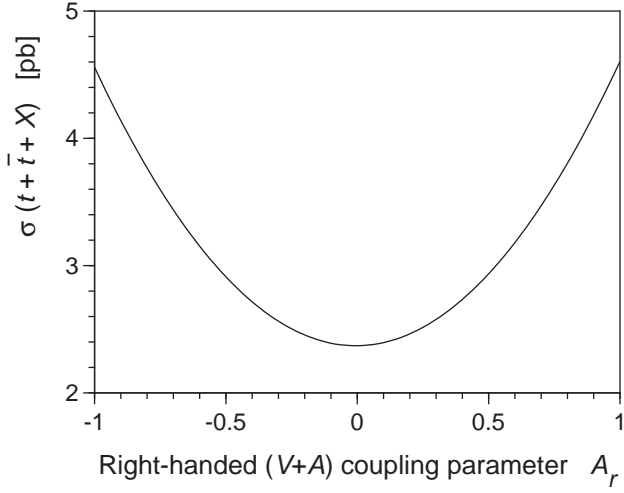}
 
\caption[fig9]{Total single top and antitop production cross section at the
upgraded Tevatron, with $\sqrt{s}=2.0$~TeV and $m_t=180$~GeV, versus the
right-handed ${\vpa}$ coupling parameter $A_r$.}

\label{Ar}
\end{figure}

The dependence of the total single top quark cross section on the parameter
$A_r$ is shown in Fig.\ \ref{Ar}, for $\sqrt{s}=2.0{\tev}$, $m_t=180{\gev}$, and
${\vtb}=0.999$. Here, $\sigma({\ppbar}{\rargap}t+{\tbar}+X)=2\times\sigma(
{\ppbar}{\rargap}t{\bbar}+tq+tq{\bbar})$. The standard model value of $A_r$ is
zero. The production rate varies almost quadratically with $A_r$, and is nearly
symmetric about the point $A_r=0$. The cross section rises from 2.44~pb when
$A_r=0$ to 4.68~pb when $A_r=-1$ and to 4.73~pb when $A_r=+1$.


\subsection{Sensitivity in the ({\vtb},$A_r$) plane}

We have calculated the region in the ({\vtb},$A_r$) plane for which there will
be experimental sensitivity using future single top measurements. If one
finds a number of single top events consistent with the standard model
prediction, then it may be that the $Wtb$ coupling is purely left-handed, and
that {\vtb} is close to unity. Alternatively, the cross section could be boosted
by an anomalous contribution to the $Wtb$ coupling, as shown for example in
Fig.\ \ref{Ar}, with {\vtb} correspondingly lower.

The error on the measurement of {\vtb} is dependent on the error on the single
top cross section, including both experimental and theoretical contributions.
First we estimate the experimental error for a top quark of mass 180{\gev} at
$\sqrt{s}=2.0{\tev}$ as follows: we take the integrated luminosity for Tevatron
Run~2 as 2~fb$^{-1}$, with an error of 5\%; the signal acceptance including at
least one $b$~tag as 0.20, from the TeV-2000 study of single top production
\cite{tev2000}, with an error of 7\%; and a signal to background ratio of 1:2,
with a systematic error on the background of 7\%. The available branching
fraction includes both the electron and muon decay channels, giving
$B=\frac{2}{9}$. In Run~2, all accessible modes of single top production will
have to be used together in order not to make a statistics-limited measurement.
We use here the value 2.44~pb for the single top cross section (= 0.72~pb
(s-channel) + 1.72~pb (t- and u-channels)) from {\sc cteq3m}. These assumptions
lead to a prediction that approximately 650 events will be found in a search,
with one third coming from single top production and two thirds from various
backgrounds (e.g.\ $W+b{\bbar}$, $W+$~light jets with a mistag, {\ttbar}).
Therefore, the experimental error on the total single top cross section will be
10\% (statistical) $\oplus$ 16\% (systematic) = 19\%, where the $\oplus$ symbol
means ``add in quadrature".

The error on the theoretical calculation of the cross section includes
contributions from the choice of parton distribution function and from the
scale, as discussed earlier in this paper, where they were found to be
$\sim$12\%. However, there is another contribution, not well quantified, from
the lack of knowledge of the gluon distribution in the proton and antiproton
for t- and u-channel single top processes. This error has been variously
reported to the authors as 30\% \cite{yuan7} and 10\% \cite{weerts}, and so we
use these values here to estimate the error on the theoretical total single top
cross section at 32\% or 16\%.

The error on a measurement of {\vtb} will be half the error on the single top
cross section, since the cross sections for all single top processes are
proportional to $|{\vtb}|^2$. This results in an error on {\vtb} of $(19\%
\oplus 32\%)/2=19\%$ or $(19\% \oplus 16\%)/2)=12\%$ from the Tevatron Run~2,
depending on one's view of the knowledge of the gluon momentum distributions in
the proton sea.

There may be a Run~3 at the Tevatron from 2002 onwards, producing 30~fb$^{-1}$
of data. This high luminosity mode of collider running is known as ``TeV33"
after the planned instantaneous luminosity of $10^{33}$~cm$^{-2}$s$^{-1}$. With
such high statistics available, it has been shown by Stelzer and Willenbrock
\cite{stelzer} that using just s-channel $W^*$ production with double
$b$~tagging instead of all single top modes with only one tag will eliminate
most of the uncertainty on the theoretical cross section, because there will no
longer be any contributions from processes with initial state gluons. They also
showed that the measurement should be possible using 3 fb$^{-1}$ of data. We
update their calculation here for Run~3, including estimates of the systematic
errors. The cross section for ${\ppbar}{\rargap}t{\bbar}+{\tbar}b+X$ is
0.716~pb, with $m_t=180{\gev}$, $\sqrt{s}=2.0{\tev}$ and $Q^2=m_t^2$. To
estimate the error on {\vtb} using Run~3 data, we make the following
assumptions: the error on the luminosity remains at 5\%; the signal acceptance
for $W^*$ single top is 0.08 when requiring a double $b$~tag, as shown in Ref.\
\cite{stelzer}, with a 1.8\% error; and the signal to background ratio is 1:2
(again from \cite{stelzer}), with a systematic error on the background of 1.8\%.
Therefore, an experiment at the Tevatron in Run~3 will see approximately 1,146
events when searching for $W^*$ single top production, with one third signal
events and two thirds coming from various backgrounds (e.g.\ $W+b{\bbar}$,
$W+$~light jets with two mistags, $WZ$ with $Z{\rar}b{\bbar}$, $W$-gluon fusion,
{\ttbar}). This observation will lead to a measurement of the $W^*$ single top
production cross section with an error of 7\% (statistical) $\oplus$ 6\%
(systematic) = 10\%. Smith and Willenbrock \cite{smith} show that the error on
the theoretical cross section for $W^*$ single top production is only 3\%,
leading to an error on {\vtb} of 5\%.

\begin{figure}
\epsfxsize=6.5in \epsfbox{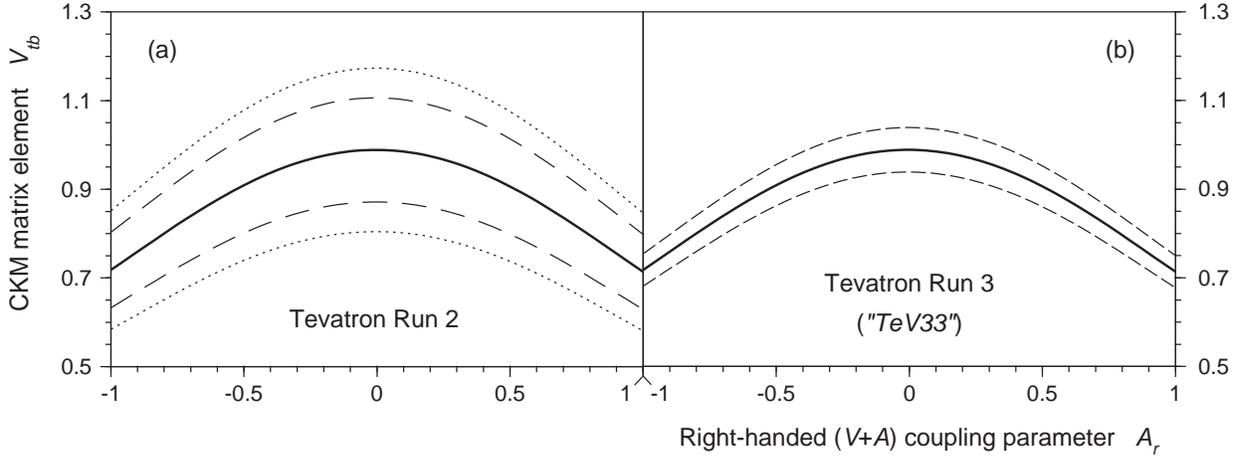}

\caption[fig10]{Estimated $1\sigma$ measurements in the ({\vtb},$A_r$) plane for
an experiment running at the upgraded Tevatron collider at $\sqrt{s}=2.0$~TeV,
assuming that the number of events seen is consistent with the standard model
prediction. Plot (a) shows the results for 2~fb$^{-1}$ of data, using all
accessible modes of single top production
(${\ppbar}{\rargap}t{\bbar}+tq+tq{\bbar}+$~c.c.). The outer short-dashed lines
enclose the region resulting from a 32\% error on the theoretical cross section
and the inner long-dashed lines from a 16\% uncertainty. Plot (b) is for a
future run with 30~fb$^{-1}$ of data, using only $W^*$ single top production
(${\ppbar}{\rargap}t{\bbar}+{\tbar}b+X$) where the error on the theoretical
cross section is 3\%.}

\label{VtbAr}
\end{figure}

In Fig.\ \ref{VtbAr} we show the results of these calculations, extended into
the ({\vtb},$A_r$) plane. In plot (a) for Tevatron Run~2 (2~fb$^{-1}$), the
outer short-dashed contours show the result when the error on the theory cross
section includes a 30\% contribution from lack of knowledge of the gluon
distribution. The inner long-dashed contours result from when this error
contributes only 10\% to the overall measurement. Plot (b) presents our
estimates for Tevatron Run~3, ``TeV33" (30~fb$^{-1}$), with the dashed contours
showing the precision obtainable using a theory error of only 3\% and an
experimental search to isolate the $W^*$ s-channel mode of single top. We
discuss in the next section how one might distinguish standard model production
from the {\vpa} scenario discussed above where the effects of an elevated cross
section caused by the anomalous coupling cancel with a reduced value of {\vtb}
from a possible mixing of the top quark with a new fourth generation quark to
give an observed number of events consistent with the standard model.


\subsection{Polarization of the top quark}

Top quark polarization depends strongly on the structure of the $Wtb$ coupling,
and one might expect an asymmetry in angular distributions of the final state
partons for different values of $A_r$. For example, standard model single top is
produced almost 100\% left-handedly polarized because of the left-handed current
structure of the $Wtb$ coupling, whereas if $A_r=1$, the top quark is not
polarized at all. To calculate polarization effects using Monte Carlo
generators, it is necessary either to keep the polarization of all particles in
the final states of the $2{\rar}2$ and $2{\rar}3$ processes being studied (i.e.\
$q'{\qbar}{\rar}t{\bbar}$, $q'b{\rar}tq$, $q'g{\rar}tq{\bbar}$), with subsequent
decays of the polarized top quark and $W$~boson ($t{\rar}Wb$, $W{\rar}e{\nu}$),
or else one needs to calculate the higher order $2{\rar}4$ and $2{\rar}5$
processes (i.e.\ $q'{\qbar}{\rar}e{\nu}b{\bbar}$, $q'b{\rar}e{\nu}bq$,
$q'g{\rar}e{\nu}bq{\bbar}$), with the top quark and $W$~boson treated as
resonances in the intermediate states. The second method automatically includes
the polarizations of the intermediate state $t$ and $W$. To study the
differences between kinematic distributions when the polarizations of the top
quark and $W$~boson have been taken into account with those where they are
assumed to be unpolarized (as in most Monte Carlo generators, e.g.\ {\sc
pythia}), we have calculated the $2{\rar}4$ and $2{\rar}5$ processes for the
three significant single top production modes using {\sc c{\small{omp}}hep}
alone, and compared the results with calculations where we used {\sc
c{\small{omp}}hep} for the $2{\rar}2$ and $2{\rar}3$ single top processes, and
{\sc pythia} for the subsequent $t$ and $W$ decays.

Our direct calculations show that the $p_T$ and $\eta$ distributions are not
sensitive to the polarization of the top quark.

Two representative examples of distributions expected to reflect the top quark
polarization effects are the invariant mass of the positron and the $b$~quark,
{\meb}, and the cosine of the polar angle, {\costh}. The invariant mass {\meb}
is given by:
\[
       m_{eb} = \sqrt{(E_e+E_b)^2 - (p_{Te}+p_{Tb})^2 - (p_{ze}+p_{zb})^2}
\]
where $p_z$ is the momentum of the positron or $b$ quark along the beam
direction.

The polar angle $\theta^*_e$ is defined as the angle between the positron
direction and the $x$ axis within the rest frame of the $W$~boson, where the $x$
axis is defined to be in the direction of motion of the $W$~boson in the rest
frame of the top quark \cite{yuan3}. The cosine of this angle is given
approximately by:
\[
       \cos\theta^*_e \simeq 1-\frac{2m_{eb}}{m_t^2-m_W^2}.
\]

\begin{figure}
\epsfxsize=6.5in \epsfbox{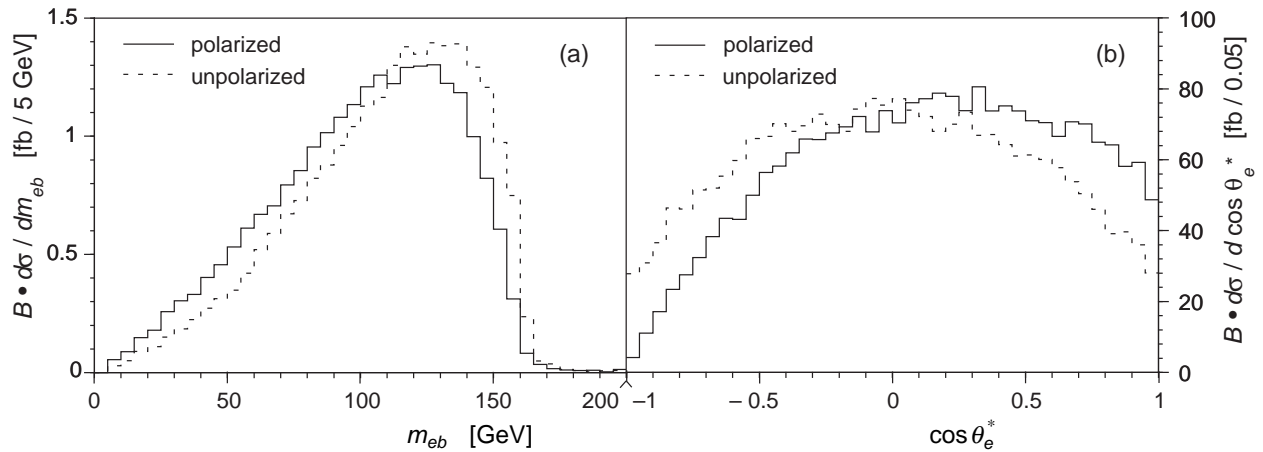}

\caption[fig11]{Single top distributions of (a) invariant mass {\meb}, and (b)
cosine of the polar angle $\theta_e^*$ (defined in the text). The solid
histograms are for the standard model case where the top quark and $W$~boson are
$\sim$100\% left-handedly polarized (fully calculated using {\sc
c{\small{omp}}hep} for the $2{\rar}4$ and $2{\rar}5$ processes with intermediate
state $t$ and $W$ resonances), and the dashed histograms are for when there is
no polarization, corresponding either to a {\vpa} term with $A_r=1$ in the $Wtb$
coupling, or to the case where the polarization has been excluded from the
calculation (e.g.\ by using {\sc pythia} to decay the $W$~boson).}

\label{polar}
\end{figure}

Figure~\ref{polar} shows the distributions of (a) {\meb} and (b) {\costh},
for the case when polarizations of the top quark and $W$~boson have
been properly taken into account (solid histogram), and for when summation over
the polarization of the top quark decay products has been done using the
subsequent decays of an unpolarized top quark and $W$~boson (dashed histogram).
One can see that there are indeed differences in these distributions for the
polarized and unpolarized cases. In particular, an asymmetry (or lack of it) in
{\costh} when the positron is emitted aligned or antialigned with the direction
of motion of the $W$~boson in the top rest frame should be observable with high
statistics. All three modes of single top production exhibit this same behavior.
The two variables {\meb} and {\costh} can also be used in combination with the
total single top production rate, which is sensitive to the $Wtb$ coupling
structure as shown previously, to further our understanding of the $Wtb$
coupling.


\subsection{Top quark partial width}

From our previous estimates of the sensitivity for measuring the single top
cross section at future Tevatron runs, we can obtain the expected precision on
the top quark partial width $\Gamma(t{\rar}Wb)$. In the standard model, the top
quark partial width is very nearly the same as its full width, since {\vtb} is
so close to 1. The top partial width is directly proportional to the single top
cross section, and so the error on the width is just the experimental
measurement error on the cross section added in quadrature with the theoretical
calculation error. Thus in Run~2 the top quark width should be measured with an
error of $19\% \oplus 32\%=37\%$ or $19\% \oplus 16\%=25\%$, depending on
whether the uncertainty on the gluon distribution function is 30\% or 10\%. In
Run~3, this precision can be improved to $10\% \oplus 3\%=10\%$, which is
comparable to what can be achieved at a linear {\ee} collider \cite{frey} using
a {\ttbar} threshold scan.


\section{Conclusions}

In this paper we have reported the results of new studies of single top quark
physics at the Fermilab Tevatron {\ppbar} collider. We have made consistent
calculations of the tree level cross sections for each mode of single top
production as a function of top quark mass, parton distribution function, QCD
scale, and collision energy. We discussed details of the calculations for
several of the subprocesses involved, and gave breakdowns of the various
contributions to the overall cross sections. For a top quark of mass 180{\gev},
at $\sqrt{s}=1.8{\tev}$, with $Q^2=m_t^2$, and taking the mean result from {\sc
cteq3m} and {\sc mrs(a$^\prime$)}, we find that the leading order total single
top plus antitop cross section is $1.76^{+0.26}_{-0.18}$~pb.

We have shown for each subprocess separately the transverse momentum and
pseudorapidity distributions of the top quark, the other quarks produced with
it, and its decay products. These kinematic distributions need to be
understood in order to be able to separate signal from background in an
experimental search.

We then considered the possibility for measuring the CKM matrix element {\vtb}
and the $Wtb$ coupling directly using single top events from the next Tevatron
run. We estimated the sensitivity such measurements might have, and how an
anomalous {\vpa} term in the $Wtb$ coupling would affect the measurement of
{\vtb}. If there is no anomalous component to the $Wtb$ coupling, then {\vtb}
can be measured to a precision of 19\% or 12\% in Run~2 (1999--2001), with the
two values coming from different estimates of the uncertainty in the gluon
distribution function. In Run~3 (2002--2006), the precision on {\vtb} will be
improved to 5\%. The top quark polarization affects the angular distributions of
its decay products, and we investigated how this could be used together with a
measurement of the single top cross section to distinguish between various
processes affecting the top quark beyond the standard model. Finally our
estimates of the single top cross section error show that the top quark partial
width will be measured to within 37\%--25\% in Run~2, and to a precision of 10\%
in Run~3.

We find the prospects for single top physics at the Tevatron exciting and
that a rich program of studies will be possible in the future.


\section*{Acknowledgements}

We would like to thank Dan~Amidei, Phil~Baringer, Pavel~Ermolov, Paul~Grannis,
Asher~Klatchko, Wu-Ki~Tung, Harry Weerts, Scott~Willenbrock, and C.-P.~Yuan for
useful discussions during this work. We would also like to thank the Moscow
State University CompHEP group, especially Slava~Ilyin and Alexander~Pukhov for
their help with the latest version of {\sc c{\small{omp}}hep}. E.B. and A.B.
thank the {\dzero} Collaboration for their kind hospitality during our stay at
Fermilab. We acknowledge the financial support of the U.S. Department of Energy
under grant number DE--FG03--94ER40837, and the Ministry of Science and
Technology Policy in Russia. This work has also been supported in part by grant
number M9B000 from the International Science Foundation, in part by RFBR grants
96-02-19773a and 96-02-18635a, and by grant 95-0-6.4-38 of the Center for
Natural Sciences of the State Committee for Higher Education in Russia.


\bibliographystyle{unsrt}

\begin{thebibliography}{99.}

\bibitem{d01}
S.~Abachi {\it et al.}, ({\dzero} Collaboration),
Phys. Rev. Lett. {\bf 74}, 2632 (1995).

\bibitem{cdf1}
F.~Abe {\it et al.}, (CDF Collaboration),
Phys. Rev. Lett. {\bf 74}, 2626 (1995).

\bibitem{grannis}
P.~Grannis, plenary talk at the International Conference on High Energy
Physics, Warsaw, 1996, reporting the analysis of the {\dzero} and CDF
collaborations.

\bibitem{ewmt}
A.~Blondel, plenary talk at the International Conference on High Energy
Physics, Warsaw, 1996, reporting the analysis of the LEP Electroweak Working
Group and the SLD Heavy Flavor Group, CERN Internal Report No. LEPEWWG/96-02,
available at http://www.cern.ch/LEPEWWG.

\bibitem{peccei}
R.D.~Peccei, S.~Peris and X.~Zhang, Nucl. Phys. {\bf B349}, 305 (1991).

\bibitem{kobayashi}
N.~Cabibbo, Phys. Rev. Lett. {\bf 10}, 531 (1963);

M.~Kobayashi and T.~Maskawa, Prog. Theor. Phys. {\bf 49}, 652 (1973).

\bibitem{pdg}
R.M.~Barnett {\it et al.}, (Particle Data Group), Phys. Rev.~D~{\bf 54}, 1
(1996).

\bibitem{dawson}
S.~Dawson, Nucl.~Phys.~{\bf B249}, 42 (1985).

\bibitem{dicus}
S.S.D.~Willenbrock and D.A.~Dicus, Phys. Rev.~D~{\bf 34}, 155 (1986).

\bibitem{willenbrock}
S.~Dawson and S.S.D.~Willenbrock, Nucl. Phys.~{\bf B284}, 449 (1987).

\bibitem{yuan2}
C.-P.~Yuan, Phys. Rev.~D~{\bf 41}, 42 (1990).

\bibitem{moers}
T.~Moers, R.~Priem, D.~Rein and H.~Reitler,
in {\it Proceedings of the ECFA Large Hadron Collider (LHC) Workshop: Physics
and Instrumentation, Aachen, Germany, 1990}, edited by G. Jarlskog and D. Rein
(CERN, 1990) vol.~2, p.~418.

\bibitem{slabosp}
G.V.~Jikia and S.R.~Slabospitsky, Phys. Lett {\bf B295}, 136 (1992).

\bibitem{parke}
R.K.~Ellis and S.~Parke, Phys. Rev.~D~{\bf 46}, 3785 (1992).

\bibitem{bordes}
G.~Bordes and B.~van~Eijk, Z. Phys. {\bf C57}, 81 (1993).

\bibitem{yuan3}
D.O.~Carlson and C.-P.~Yuan, Phys. Lett. {\bf B306}, 386 (1993).

\bibitem{bordes2}
G.~Bordes and B.~van~Eijk, Nucl. Phys. {\bf B435}, 23 (1995).

\bibitem{yuan3a}
D.O.~Carlson and C.-P.~Yuan, MSUHEP--40903 (March 1995).

\bibitem{yuan3b}
D.O.~Carlson and C.-P.~Yuan, to appear in {\it Proceedings of the Workshop on
Physics of the Top Quark, Ames, Iowa}, (May 1995), hep-ph/9509208.

\bibitem{heinson}
A.P.~Heinson, A.S.~Belyaev and E.E.~Boos, to appear in {\it Proceedings of the
Workshop on Physics of the Top Quark, Ames, Iowa}, (May 1995), hep-ph/9509274.

\bibitem{cortese}
S.~Cortese and R.~Petronzio, Phys. Lett. {\bf B253}, 494 (1991).

\bibitem{stelzer}
T.~Stelzer and S.~Willenbrock, Phys. Lett. {\bf B357}, 125 (1995).

\bibitem{pittau}
R.~Pittau, Phys. Lett. {\bf B386}, 397 (1996).

\bibitem{smith}
M.~Smith and S. Willenbrock, Phys. Rev.~D~{\bf 54}, 6696 (1996).

\bibitem{atwoodbar}
D.~Atwood, S.~Bar-Shalom, G.~Eilam and A.~Soni, Phys. Rev. D {\bf 54}, 5412
(1996).

\bibitem{li1}
C.S.~Li, R.J.~Oakes and J.M.~Yang, to appear in Phys. Rev. D, hep-ph/9608460.

\bibitem{li2}
C.S.~Li, R.J.~Oakes and J.M.~Yang, hep-ph/9611455.

\bibitem{yuan4}
D.O.~Carlson, E.~Malkawi and C.-P.~Yuan, Phys. Lett. {\bf B337}, 145 (1994).

\bibitem{yuan5}
E.~Malkawi and C.-P.~Yuan, Phys. Rev.~D~{\bf 50}, 4462 (1994).

\bibitem{yuan6}
C.-P.~Yuan, Mod. Phys. Lett. {\bf A10}, 627 (1995).

\bibitem{rizzo}
T.G.~Rizzo, Phys. Rev. D {\bf 53}, 6218 (1996).

\bibitem{mahlon}
G.~Mahlon and S.~Parke, hep-ph/9611367.

\bibitem{katuya}
M.~Katuya, J.~Morishita, T.~Munehisa and Y.~Shimizu, Prog. Theor. Phys.
{\bf 75}, 92 (1986).

\bibitem{barger}
V.~Barger and K.~Hagiwara, Phys. Rev.~D~{\bf 37}, 3320 (1988).

\bibitem{raidal}
M.~Raidal and R.~Vuopionper\"{a}, Phys. Lett. {\bf B318}, 237 (1993).

\bibitem{panella}
O.~Panella, G.~Pancheri and Y.N.~Srivastara,
Phys. Lett. {\bf B318}, 214 (1993).

\bibitem{hagiwara}
K.~Hagiwara, M.~Tanaka and T.~Stelzer, Phys. Lett. {\bf B325}, 521 (1994).

\bibitem{mele}
S.~Ambrosanio and B.~Mele, Z. Phys. {\bf C63}, 63 (1994).

\bibitem{boos}
E.~Boos {\it et al.}, Phys. Lett. {\bf B326}, 190 (1994).

\bibitem{jikia1}
N.V.~Dokholian and G.V.~Jikia, Phys. Lett. {\bf B336}, 251 (1994).

\bibitem{booskuri}
E.~Boos, Y.~Kurihara, Y.~Shimizu, M.~Sachwitz, H.J.~Schreiber and
S.~Shichanin, Z.~Phys. {\bf C70}, 255 (1996).

\bibitem{qian}
M.-L.~Qian, W.-G.~Ma, L.-Z.~Sun and Y.-Y.~Liu, J. Phys. {\bf G20}, 895 (1994).

\bibitem{ballestrero}
A.~Ballestrero, V.A.~Khoze, E.~Maina, S.~Moretti and W.J.~Stirling,
Z.~Phys. {\bf C72}, 71 (1996).

\bibitem{jikia2}
G.V.~Jikia, Nucl. Phys. {\bf B374}, 83 (1992).

\bibitem{yehudai}
E.~Yehudai, S.~Godfrey and K.A.~Peterson,
in {\it Proceedings of the Second International Workshop on Physics and
Experiments with Linear {\ee} Colliders, Waikoloa, Hawaii, 1993},
edited by F.A. Harris, S.L. Olsen, S. Pakvasa and X. Tata, (World Scientific,
1993) vol.~2, p.~568.

\bibitem{pukhov}
E.~Boos, A.~Pukhov, M.~Sachwitz and H.J. Shreiber, to appear in Z. Phys. {\bf
C}, hep-ph/9610424.

\bibitem{comphep}
E.E.~Boos {\it et al.}, in {\it Proceedings of the XXVIth Rencontres de
Moriond, High Energy Hadronic Interactions, Les Arcs, France, 1991}, edited by
J.~Tran~Than~Van, (Editions Frontieres, 1991) p.~501;

E.E.~Boos {\it et al.}, in {\it Proceedings of the Second International
Workshop on Software Engineering, Artificial Intelligence and Expert Systems
for High Energy and Nuclear Physics, La Londe Les Maures, France, 1992}, edited
by D.~Perret-Gallix, (World Scientific, 1992), p.~665.

{\sc c{\small{omp}}hep} is available at
http://theory.npi.msu.su/$\sim$pukhov/comphep.html

\bibitem{bases}
S.~Kawabata, Comp. Phys. Commun. {\bf 41}, 127 (1986).

\bibitem{kovalenko}
V.A.~Ilyin, D.N.~Kovalenko and A.E.~Pukhov, INP MSU Preprint-95-2/366, Moscow
State University, (1995).

\bibitem{vegas}
G.P.~Lepage, J. Comput. Phys. {\bf 27}, 192 (1978).

\bibitem{pythia}
T.~Sj\"{o}strand, Comput. Phys. Commun. {\bf 82}, 74 (1994).
P{\sc ythia} version 5.7 was used.

\bibitem{cteq}
H.L.~Lai {\it et al.}, (CTEQ Collaboration), Phys. Rev.~D~{\bf 51}, 4763 (1995).

\bibitem{mrs}
A.D.~Martin, R.G.~Roberts, W.J.~Stirling, Phys. Lett. {\bf B354}, 155 (1995).

\bibitem{bardeen}
W.A.~Bardeen, A.J.~Buras, D.W.~Duke and T.~Muta, Phys. Rev.~D~{\bf 18}, 3998
(1978).

\bibitem{cteq4}
H.L.~Lai {\it et al.}, (CTEQ Collaboration), hep-ph/9606399.

\bibitem{mrsr}
A.D.~Martin, R.G.~Roberts and W.J.~Stirling, Phys. Lett. {\bf B387}, 419 (1996).

\bibitem{dglap}
V.N.~Gribov and L.N.~Lipatov, Sov. J. Nucl. Phys. {\bf 15}, 438 (1972);
{\bf 15}, 675 (1972);

Yu.L.~Dokshitzer, Sov. Phys. JETP {\bf 46}, 641 (1977);

G.~Altarelli and G.~Parisi, Nucl. Phys. {\bf B126}, 298 (1977).

\bibitem{barnett}
R.M.~Barnett, H.E.~Haber and D.E.~Soper, Nucl. Phys. {\bf B306}, 697 (1988).

\bibitem{olness}
F.I.~Olness and W.-K.~Tung, Nucl. Phys. {\bf B308}, 813 (1988).

\bibitem{tung}
W.-K.~Tung, private communication.

\bibitem{berger}
E.L.~Berger and H.~Contopanagos, Phys. Lett. {\bf B361}, 115 (1995);
Phys. Rev. D {\bf 54}, 3085 (1996).

\bibitem{cteq2}
J.~Botts {\it et al.}, (CTEQ Collaboration), Phys. Lett. {\bf B304}, 159 (1993).

\bibitem{tev2000a}
U.~Heintz {\it et al.}, ``Light Higgs Physics at the Tevatron", in
{\it Future Electroweak Physics at the Fermilab Tevatron -- Report of the
tev\_2000 Study Group, Fermilab, 1996}, edited by D.~Amidei and R.~Brock,
(FERMILAB--PUB--96/082), p.~136.

\bibitem{herwig}
G.~Marchesini {\it et al.}, Comp. Phys. Commun. {\bf 67}, 465 (1992).

\bibitem{deshpande}
N.G.~Deshpande, B.~Margolis and H.D. Trottier, Phys. Rev. D {\bf 45}, 178
(1992).

\bibitem{atwood}
D.~Atwood, A.~Aeppli and A.~Soni, Phys. Rev. Lett. {\bf 69}, 2754 (1992).

\bibitem{atwood2}
D.~Atwood, A.~Kagan and T.G.~Rizzo, Phys. Rev. D {\bf52}, 6264 (1995).

\bibitem{huang}
C.-S.~Huang and T.-J.~Li, Z. Phys. {\bf C68}, 319 (1995).

\bibitem{cheung}
K.~Cheung, Phys. Rev. D {\bf 53}, 3604 (1996).

\bibitem{jberger}
J.~Berger, A.~Blotz, H.-C.~Kim and K.~Goeke, Phys. Rev. D {\bf 54}, 3598 (1996).

\bibitem{tait}
T.~Tait and C.-P.~Yuan, hep-ph/9611244.

\bibitem{hill}
C.T.~Hill, Phys. Lett. {\bf B266}, 419 (1991).

\bibitem{hill2}
C.T.~Hill and S.J.~Parke, Phys. Rev. D {\bf 49}, 4454 (1994).

\bibitem{eichten}
E.~Eichten and K.~Lane, Phys. Lett. {\bf B327}, 129 (1994).

\bibitem{lsmith}
L.L.~Smith, P.~Jain and D.W.~McKay, to appear in {\it Proceedings of the 9th
Annual American Physical Society Division of Particles and Fields Meeting,
Minneapolis, MN,} (August 1996), hep-ph/9608328.

\bibitem{datta}
A.~Datta and X.~Xhang, hep-ph/9611247.

\bibitem{simmons}
E.H.~Simmons, hep-ph/9612402.

\bibitem{tev2000}
D.~Amidei {\it et al.}, ``Top Physics", in {\it Future Electroweak Physics
at the Fermilab Tevatron -- Report of the tev\_2000 Study Group, Fermilab,
1996}, edited by D.~Amidei and R.~Brock, (FERMILAB--PUB--96/082), p.~13.

\bibitem{yuan7}
C.-P.~Yuan, private communication.

\bibitem{weerts}
H.~Weerts, private communication.

\bibitem{frey}
R.~Frey, to appear in {\it Proceedings of the 3rd Workshop on Physics and
Experiments with {\ee} Linear Colliders, Iwate, Japan, 1995},
hep-ph/9606201.

\end{thebibliography}


\twocolumn

\begin{table}

\begin{tabular}{lcc}
 Single Top Process & \multicolumn{2}{c} {Cross Section [pb]}        \\
 & {\sc cteq3m} & {\sc mrs(a$^\prime$)}                              \\
\hline
 \bf{1.}~${\bfppbar}{\bfrargap}{\bftbbar}\bf{+X}$ & \bf{0.2847} & \bf{0.2772}
\medskip \\
 ~~~~\bf{1.1}~~${\bfqpqbar}{\bfrar}{\bftbbar}$
                                         & \bf{0.2510} & \bf{0.2452} \\
 ~~~~~~~~~~$u{\dbar}{\rar}t{\bbar}$                & 0.2423 & 0.2370 \\
 ~~~~~~~~~~${\dbar}u{\rar}t{\bbar}$                & 0.0044 & 0.0040 \\
 ~~~~~~~~~~Other modes                             & 0.0043 & 0.0042
\smallskip \\
 ~~~~\bf{1.2}~~$\bf{q'g}{\bfrar}{\bftbbarq}$
                                         & \bf{0.0337} & \bf{0.0320} \\
 ~~~~~~~~~~$ug{\rar}t{\bbar}d$                     & 0.0213 & 0.0200 \\
 ~~~~~~~~~~$gu{\rar}t{\bbar}d$                     & 0.0009 & 0.0009 \\
 ~~~~~~~~~~${\dbar}g{\rar}t{\bbar}{\ubar}$         & 0.0016 & 0.0015 \\
 ~~~~~~~~~~$g{\dbar}{\rar}t{\bbar}{\ubar}$         & 0.0080 & 0.0077 \\
 ~~~~~~~~~~Other modes                             & 0.0019 & 0.0019
\bigskip \\
 \bf{2.}~${\bfppbar}{\bfrargap}\bf{tq+X}$  & \bf{0.5697} & \bf{0.5059}
\medskip \\
 ~~~~\bf{2.1}~~$\bf{q'b}{\bfrar}\bf{tq}$ & \bf{0.2448} & \bf{0.2013}
\smallskip \\
 ~~~~~~~~~~$ub{\rar}td$                            & 0.1607 & 0.1333 \\
 ~~~~~~~~~~$bu{\rar}td$                            & 0.0063 & 0.0048 \\
 ~~~~~~~~~~${\dbar}b{\rar}t{\ubar}$                & 0.0075 & 0.0056 \\
 ~~~~~~~~~~$b{\dbar}{\rar}t{\ubar}$                & 0.0403 & 0.0338 \\
 ~~~~~~~~~~Other modes                             & 0.0300 & 0.0238
\smallskip \\
 ~~~~\bf{2.2}~~$\bf{q'g}{\bfrar}{\bftqbbar}$
                                         & \bf{0.3249} & \bf{0.3046} \\
 ~~~~~~~~~~$ug{\rar}td{\bbar}$                     & 0.2162 & 0.2034 \\
 ~~~~~~~~~~$gu{\rar}td{\bbar}$                     & 0.0080 & 0.0073 \\
 ~~~~~~~~~~${\dbar}g{\rar}t{\ubar}{\bbar}$         & 0.0100 & 0.0089 \\
 ~~~~~~~~~~$g{\dbar}{\rar}t{\ubar}{\bbar}$         & 0.0559 & 0.0540 \\
 ~~~~~~~~~~Other modes                             & 0.0348 & 0.0310
\bigskip \\
 \bf{3.}~${\bfppbar}{\bfrargap}\bf{tW+X}$  & \bf{0.0658} & \bf{0.0573}
\medskip \\
 ~~~~\bf{3.1}~~$\bf{bg}{\bfrar}\bf{tW}$
                                         & \bf{0.0418} & \bf{0.0346} \\
 ~~~~~~~~~~$bg{\rar}tW$                            & 0.0209 & 0.0173 \\
 ~~~~~~~~~~$gb{\rar}tW$                            & 0.0209 & 0.0173
\smallskip \\
 ~~~~\bf{3.2}~~${\bfqqbar}{\bfrar}{\bftWbbar}$
                                         & \bf{0.0027} & \bf{0.0026} \\
 ~~~~~~~~~~$u{\ubar}{\rar}tW{\bbar}$               & 0.0024 & 0.0023 \\
 ~~~~~~~~~~${\ubar}u{\rar}tW{\bbar}$               & 0.0000 & 0.0000 \\
 ~~~~~~~~~~$d{\dbar}{\rar}tW{\bbar}$               & 0.0003 & 0.0003 \\
 ~~~~~~~~~~${\dbar}d{\rar}tW{\bbar}$               & 0.0000 & 0.0000
\smallskip \\
 ~~~~\bf{3.3}~~$\bf{gg}{\bfrar}{\bftWbbar}$
                                         & \bf{0.0213} & \bf{0.0201}
\bigskip \\
 $\bf{\sigma(}{\bfppbar}{\bfrargap}\bf{t+X)}$ & \bf{0.9202} & \bf{0.8404}
\smallskip \\
\end{tabular}

\caption[tab1]{Production cross sections for single top quark processes for
$m_t=180$~GeV, with $Q^2=m_t^2$ and $\sqrt{s}=1.8$~TeV, using $\sc {cteq3m}$
and $\sc {mrs(a^\prime)}$ parton distributions. ``Other modes" refers
to subprocesses with an $s$ or $c$ quark in the initial state, and to
subprocesses involving an off-diagonal CKM matrix element term.}

\label{table1}
\end{table}


\onecolumn
\begin{table}

\begin{tabular}{cccccccccc}
 \multicolumn{4}{r}
 {Total Single Top Cross Section [pb]} & & & & &
\medskip \\
 & \multicolumn{3}{c}
{$\bf{\sigma(}{\bfppbar}{\bfrar}\bf{t+}{\bftbar}\bf{+X)}$}
 & \multicolumn{3}{c}
{$\bf{1.}~\bf{\sigma(}{\bfppbar}{\bfrar}{\bftbbar}\bf{+}{\bftbarb}\bf{+X)}$}
 & \multicolumn{3}{c}
{$\bf{3.}~\bf{\sigma(}{\bfppbar}{\bfrar}\bf{tW+}{\bftbarW}\bf{+X)}$}
\smallskip \\
  $m_t$  & Lower & Central & Upper
         & Lower & Central & Upper
         & Lower & Central & Upper \\
 {[GeV]} & bound &  value  & bound
         & bound &  value  & bound
         & bound &  value  & bound
\smallskip \\
\hline
 140 & 4.15 & 4.61 & 5.19 & 1.70 & 1.83 & 2.02 & 0.32 & 0.39 & 0.48 \\
 150 & 3.20 & 3.56 & 4.03 & 1.21 & 1.33 & 1.45 & 0.24 & 0.29 & 0.36 \\
 160 & 2.51 & 2.79 & 3.17 & 0.88 & 0.98 & 1.07 & 0.18 & 0.22 & 0.28 \\
 170 & 1.98 & 2.21 & 2.52 & 0.66 & 0.74 & 0.80 & 0.14 & 0.16 & 0.21 \\
 180 & 1.58 & 1.76 & 2.02 & 0.50 & 0.56 & 0.61 & 0.10 & 0.12 & 0.16 \\
 190 & 1.26 & 1.42 & 1.63 & 0.39 & 0.43 & 0.47 & 0.08 & 0.09 & 0.12 \\
 200 & 1.02 & 1.15 & 1.32 & 0.30 & 0.34 & 0.37 & 0.06 & 0.07 & 0.09 \\
 210 & 0.83 & 0.93 & 1.07 & 0.24 & 0.27 & 0.30 & 0.04 & 0.06 & 0.07 \\
 220 & 0.67 & 0.76 & 0.88 & 0.19 & 0.21 & 0.24 & 0.03 & 0.04 & 0.05 \\
\hline
\hline
 & \multicolumn{3}{l}
{$\bf{2.1}~~\bf{\sigma(}\bf{q'b}{\bfrar}\bf{tq},
                                       {\bfqpbarbbar}{\bfrar}{\bftbarqbar})$}
 & \multicolumn{3}{l}
{$\bf{2.2}~~\bf{\sigma(}\bf{q'g}{\bfrar}{\bftqbbar},
                                       {\bfqpbarg}{\bfrar}{\bftbarqbarb})$}
 & \multicolumn{3}{c}
{$\bf{2.}~\bf{\sigma(}{\bfppbar}{\bfrar}\bf{tq+}{\bftbarqbar}\bf{+X)}$}
\smallskip \\
  $m_t$  & Lower & Central & Upper
         & Lower & Central & Upper
         & Lower & Central & Upper \\
 {[GeV]} & bound &  value  & bound
         & bound &  value  & bound
         & bound &  value  & bound
\smallskip \\
\hline
 140 & 0.85 & 0.95 & 1.07 & 1.13 & 1.43 & 1.81 & 2.13 & 2.38 & 2.69 \\
 150 & 0.69 & 0.78 & 0.88 & 0.92 & 1.16 & 1.49 & 1.75 & 1.94 & 2.22 \\
 160 & 0.57 & 0.65 & 0.73 & 0.75 & 0.94 & 1.23 & 1.44 & 1.59 & 1.83 \\
 170 & 0.47 & 0.54 & 0.60 & 0.62 & 0.77 & 1.01 & 1.18 & 1.31 & 1.51 \\
 180 & 0.39 & 0.45 & 0.50 & 0.50 & 0.63 & 0.83 & 0.97 & 1.08 & 1.25 \\
 190 & 0.32 & 0.37 & 0.42 & 0.41 & 0.52 & 0.68 & 0.80 & 0.89 & 1.03 \\
 200 & 0.27 & 0.31 & 0.35 & 0.34 & 0.43 & 0.56 & 0.66 & 0.74 & 0.86 \\
 210 & 0.23 & 0.26 & 0.29 & 0.28 & 0.35 & 0.46 & 0.54 & 0.61 & 0.71 \\
 220 & 0.19 & 0.22 & 0.25 & 0.22 & 0.29 & 0.38 & 0.44 & 0.51 & 0.58 \\
\end{tabular}

\vspace{0.375 in}

\caption[tab2]{Single top plus antitop cross sections at the Tevatron with
$\sqrt{s}=1.8$~TeV and $Q^2=m_t^2$, as a function of top quark mass. The central
values are the mean of the calculations using {\sc cteq3m} and {\sc
mrs(a$^\prime$)}. The upper and lower bounds include the effects from choice of
scale $Q^2$ and half the difference between the parton distribution functions.}

\label{table2}
\end{table}


\vspace{0.5 in}

\vspace{1 in}

\vspace{0.5 in}

\end{document}